
 \input phyzzx
\Pubnum={\vbox{ \hbox{CERN-TH.6687/92}\hbox{EHU-FT-92/5}}}
\pubnum={CERN-TH.6687/92}
\date={October, 1992}
\pubtype={}
\titlepage

\title{DOUBLE SCALING LIMIT OF \break THE SUPERVIRASORO CONSTRAINTS}
\vskip 1.0cm
\author{L. Alvarez-Gaum\'e \break K.
Becker\foot{$^{\dagger}$Permanent  address: Universit\"at Bonn,
Physikalisches Institut, Nussallee 12, W-5300 Bonn 1, Germany.}
\break and \break  M. Becker$^{\dagger}$} \address{Theory Division,
CERN\break
 CH-1211 Geneva 23, Switzerland\break}
\author{R. Emparan\break and \break J. Ma\~nes}\address{Departamento
de F\'\i sica, UPV\break Facultad de Ciencias, Bilbao, Spain}
\abstract{We obtain the double scaling limit of a set of
superloop equations recently proposed to describe the coupling
of two-dimensional supergravity to minimal superconformal
matter of type $(2,4m)$. The continuum loop equations are
described in terms of a ${\widehat c}=1$ theory with a
$Z_2$-twisted scalar field and a Weyl-Majorana fermion in the
Ramond sector. We have computed correlation functions in genus
zero, one and partially in genus two. An integrable supersymmetric
hierarchy describing our model has not yet been found. We
present a heuristic argument showing that the purely
bosonic part of our model is described by the
KdV-hierarchy.}

\endpage
\pagenumber=1

\def\rsloops{L. Alvarez-Gaum\'e, H. Itoyama, J. Ma\~nes and A.
Zadra, {\it Superloop Equations and Two-Dimensional
Supergravity}. CERN-TH.6329/91. {\it Int. J. Mod. Phys.}, to
appear.}

\def\rfqs{ D. Friedan, Z. Qiu and
 S. Shenker, {\it Phys. Rev. Lett.} {\bf 51} (1984) 1575.}

\def\rkpz{ V.G. Knizhnik, A.M. Polyakov
 and A.B. Zamolodchikov, {\it Mod. Phys. Lett.} {\bf
A3} (1988)  819.}

\def\rmatmod{ F. David, {\it Nucl. Phys.} {\bf B257} (1985) 45,
 V. Kazakov, {\it Phys. Lett.} {\bf 150B} (1985) 282,
 V. Kazakov, {\it Phys. Lett.} {\bf 119B} (1986) 140,
 I.K. Kostov and M.L. Mehta,
{\it Phys. Lett.} {\bf 189B} (1987)118.}

\def\rpz{ A.M. Polyakov and
A.B. Zamolodchikov, {\it Mod Phys. Lett.}
{\bf A3} (1988) 1213.}

\def\rdhk{J. Distler, Z. Hlousek and H. Kawai,
{\it Int. J. Mod. Phys.} {\bf A5} (1990) 391.}

\def\rdoubles{ E. Br\'ezin and V.A. Kazakov,
{\it Phys. Lett.} {\bf 236B} (1990) 144;
 M.R. Douglas and S.H. Shenker,
{\it Nucl. Phys.} {\bf B335} (1990) 635;
 D.J. Gross and A.A. Migdal,
{\it Phys. Rev. Lett.} {\bf 64} (1990) 127.
D.J. Gross and A. A. Migdal,
{\it Nucl Phys.} {\bf B340} (1990) 333.}

\def\rdouglas{ M.R. Douglas,
{\it Phys. Lett.} {\bf 238B} (1990) 176.}

\def\rdifdk{ P. Di Francesco,
J. Distler and D. Kutasov,
{\it Mod. Phys. Lett.}{\bf A5} (1990) 2135.}

\def\rmr{ Yu.I. Manin and A.O. Radul,
{\it Commun. Math. Phys.} {\bf 98} (1985) 65.}

\def\rdvv{ R. Dijkgraaf, E. Verlinde and
H. Verlinde, {\it Nucl. Phys.} {\bf B328} (1991) 435.}

\def\rfkn{ M. Fukuma, H. Kawai and
R. Nakayama,{\it Int. J. Mod Phys.} {\bf A6} (1991) 1385.}

\def\rkazakov{ V.A. Kazakov, {\it Mod.
Phys. Lett.} {\bf A4} (1989) 2125.}

\def\rsliouv{ E. Abdalla, M.C.B. Abdalla,
D. Dalmazi and K. Harada, IFT preprint
 IFT-91-0351;
 K. Aoki and E. D'Hoker, UCLA preprint UCLA-91-TEP-33;
 L. Alvarez-Gaum\'e and P. Zaugg,
 CERN preprint CERN-TH-6242-91.}

\def\rwitkon{ E. Witten,``Two Dimensional
Gravity and Intersection Theory on Moduli
 Space", {\it Surveys in Diff. Geom.}
{\bf 1} (1991) 243, and references therein,
 R. Dijkgraaf and E. Witten, {\it Nucl. Phys.}
{\bf B342} (1990) 486;
 M. Kontsevich, ``Intersection Theory
on the Moduli Space of Curves and the
 Matrix Airy Function", Bonn preprint MPI/91-47;
E. Witten, ``On the Kontsevich Model
and Other Models of Two Dimensional
 Gravity", IAS preprint, HEP-91/24.}

\def\rrabin{ J.M. Rabin, {\it Commun.
Math. Phys.} {\bf 137} (1991) 533.}

\def\rmigdal{ For details and references on
the loop equations see A.A. Migdal, {\it Phys.
Rep.} {\bf 102} (1983) 199.}

\def\rmmreview{V. Kazakov, ``Bosonic Strings
and String Field Theories in One-Dimensional
 Target Space", lecture given at
Carg\`ese, France, May 1990, to appear in
``Random Surfaces and Quantum gravity", ed.
by O. Alvarez et.al.;
  L. Alvarez-Gaum\'e,
{\it Helv. Phys. Acta} {\bf 64} (1991) 359, and references therein.}

\def\rskdv{B.A. Kuperschmidt, {\it Phys. Lett.}{\bf 102A}
(1984) 213; M. Chaichan and P.P. Kulish, {\it Phys. Lett.}
{\bf 183B} 169;  P. Mathieu, {\it Phys. Lett.}{\bf 203B}
(1988) 287, {\it Lett. Math. Phys.} {\bf 16} (1988) 277,
{\it J. Math. Phys.}{\bf 29} (1988) 287;
A. Bilal and J.-L. Gervais, {\it Phys. Lett.}
{\bf 211B} (1988) 95.}

\def\ram{J. Ambjorn and Yu. M. Makeenko, {\it Mod. Phys. Lett.}
{\bf A5} (1990) 1753;
J. Ambjorn, J. Jurkiewicz and Yu. M. Makeenko,
{\it Phys. Lett.} {\bf 251B} (1990) 517.}


\def\D{\Delta}
\def\DE{{(p-x)(p-y)}}
\def\INT{{V'(p)\over \sqrt{\DE}}}
\def\V{V'(p)}
\def\p{\partial}

\def\L{\Lambda}

\def\nkpp{n_{k+1}^{+}}
\def\nkm{n_k^{-}}

\def\k{\kappa^2}
\def\v{{\widehat v}}

\def\vo{{\widehat v_0}}

\def\z{z+u}
\def\l{\lambda}
\def\DLR{{\buildrel \leftrightarrow \over \p}}
\def\t{\tau(u)}
\def\DDLR{{\buildrel \leftrightarrow \over D}}
\def\n{n_0^{+}}
\def\px{\p _x}
\def\py{\p _y}
\def\pxy{\p ^2_{xy}}
\def\pp{\p _{+}}
\def\pms{\p _{-}}
\def\ppm{\p _{+-}^2}
\def\pt{\p _t}
\def\up{u_{+}}
\def\um{u_{-}}
\def\zp{z+\up}
\def\pint{\oint _\infty {dp\over 2\pi i}}
\def\qint{\oint _\infty {dq\over 2\pi i}}
\def\u{{\widehat u}}
\def\uo{{\widehat u_0}}
\def\Nt{{\widetilde N}}
\def\h{{1\over 2}}
\def\t{\theta}
\def\sz{\sqrt{\zp}}
\def\tz{(\zp )^{3/2}}
\def\tp{{\tilde t}^{+}}
\def\tm{{\tilde t}^{-}}
\def\Tp{{\tilde \tau}^{+}}
\def\Tm{{\tilde \tau}^{-}}
\def\to{\rightarrow}

\def\eqcxiv{
$$
u_0(p)=C(x,y)+{\L (x,y)\over p}+O({1\over p^2})
\eqn\cxiv
$$}

\def\eqcxv{
$$\eqalign{
{u_0\over\sqrt\DE}&=\INT -M(p)\cr
&={C(x,y)\over p}+{\L (x,y)+{x+y\over
2}C(x,y)\over p^2}+O({1\over p^3})\cr
}\eqn\cxv
$$}

\def\eqcxvi{
$$
C(x,y)=-\pint \INT
\eqn\cxvi
$$}

\def\eqcxvii{
$$
\L (x,y)=-\pint\INT (p-{x+y\over 2})
\eqn\cxvii
$$}

\def\eqcxviia{
$$
C(x,y)=0
\eqn\cxviia
$$}

\def\eqcxviii{
$$
M(p)=\sum _{n\geq 0}M_np^n
\eqn\cxviii
$$}

\def\eqcxix{
$$
M_n=-\qint {q^{-n-1}V'(q)\over\sqrt{(q-x)(q-y)}}
\eqn\cxix
$$}

\def\eqcxx{
$$
M(p)=-\qint {V'(q)\over (q-p)\sqrt{(q-x)(q-y)}}
\eqn\cxx
$$}

\def\eqcxxa{
$$
\px\L={1\over 4}(x-y)M(x)\;\; ,\qquad\py\L =
{1\over 4}(y-x)M(y)
\eqn\cxxa
$$}

\def\eqcxxb{
$$\eqalign{
&M(x)=m_0^- +(x-y)m_0^+\cr
&M(y)=m_0^- +(y-x)m_0^+\cr}
\eqn\cxxb
$$}

\def\eqcxxc{
$$\eqalign{
&m_0^-={2\over (x-y)}(\px -\py )\L\cr
&m_0^+={2\over (x-y)^2}(\px +\py )\L\cr}
\eqn\cxxc
$$}

\def\eqcxxvi{
$$\eqalign{
du_0(p)&={1\over\sqrt{\D}}\lbrack -\h M(p)d\D -\D dM(p)\rbrack\cr
&={d\L\over p}+O({1\over p^2})\cr}
\eqn\cxxvi
$$}

\def\eqcxxvii{
$$
d\L =\h M(p)d\D -\D dM(p)
\eqn\cxxvii
$$}

\def\eqcxxix{
$$
\uo (p)\equiv u_0(p)-\V
\eqn\cxxix
$$}

\def\eqcxxx{
$$
{\p\uo (p)\over\p\L}={1\over\sqrt\DE}
\eqn\cxxx
$$}

\def\eqcxxxiv{
$$
\Nt\equiv\sum_{k\geq 0}(\nkm+\D '\nkpp )\D ^k
\eqn\cxxxiv
$$}

\def\eqcxxxv{
$$
v(p)+{\D '\over\sqrt{\D}}\n =\xi (p)-\Nt\sqrt{\D}
\eqn\cxxxv
$$}

\def\eqcxxxva{
$$
\n =-\h\pint{\xi (p)\over\sqrt\DE}
\eqn\cxxxva
$$}

\def\eqcxxxvb{
$$
\Nt (p)=-\qint{\xi (q)\over (q-p)\sqrt{(q-x)(q-y)}}
\eqn\cxxxvb
$$}

\def\eqcxxxvi{
$$
\px \bigl( v(p)+{\D '\over\sqrt{\D}}\n )=
{1\over\sqrt{\D}}(\h (p-y)\Nt-\D\px\Nt \bigr)
\eqn\cxxxvi
$$}

\def\eqcxxxvii{
$$
\px (v(p)+{\D '\over\sqrt{\D}}\n )=2\px\n +O({1\over p})
\eqn\cxxxvii
$$}

\def\eqcxxxviii{
$$
\h (p-y)\Nt -\D\px\Nt =a+b\D '
\eqn\cxxxviii
$$}

\def\eqcxl{
$$
{1\over\sqrt{\D}} (a+b\D ')=2b+O({1\over p})
\eqn\cxl
$$}

\def\eqcxlii{
$$
\px (v(p)+{\D '\over\sqrt{\D}}\n )={2(p-y)\px\n\over\D^ \h}
\eqn\cxlii
$$}

\def\eqcxliia{
$$
\h (p-y)\Nt -\D\px\Nt =2(p-y)\px\n
\eqn\cxliia
$$}

\def\eqcxliib{
$$
\h (n_0^- +(x-y)n_1^+)=2\px\n
\eqn\cxliib
$$}

\def\eqcxliibb{
$$
\h (n_0^- +(y-x)n_1^+)=2\py\n
\eqn\cxliibb
$$}

\def\eqcxliic{
$$\eqalign{
&n_0^-=2(\px +\py )\n\cr
&n_1^+={2\over x-y}(\px -\py )\n\cr
}\eqn\cxliic
$$}

\def\eqcxliii{
$$
\py (v(p)+{\D '\over\sqrt{\D}}\n )={2(p-x)\py\n\over\D^ \h}
\eqn\cxliii
$$}

\def\eqcxlv{
$$
\u (l)\equiv
\lim_{n\rightarrow\infty}\langle\sum_{i=1}^N
\l_i^n\rangle =\lim_{n\rightarrow\infty}{N\over
\L}u^{(n)}  \eqn\cxlv
$$}

\def\eqcxlvi{
$$
\p_\L\uo (l)=\lim_{n\rightarrow\infty}{1\over 2\pi i}
{N\over\L}\oint_C{dp\;
p^n\over\sqrt\DE}
\eqn\cxlvi
$$}

\def\eqcxlvii{
$$
{1\over 2\pi i}\oint_C{dp\; p^n\over\sqrt\DE} =
{1\over 2\pi}\oint_C{d\t
e^{ila^{-2/m}\t}\over\sqrt{(e^{i\t}-x)(e^{i\t}-y)}}
\eqn\cxlvii
$$}

\def\eqcxlviib{
$$
N=a^{-2-1/m}(x_c-y_c)^\h{1\over\kappa}
\eqn\cxlviib
$$}

\def\eqcxlviii{
$$
-\pt\uo (l)=\lim_{a\rightarrow 0}\;
a^{1/m}{\sqrt{(x_c-y_c)}\over\kappa}{1\over
2\pi}\oint_C{dze^{ilz}
\over\sqrt{(e^{ia^{2/m}z}-x)(e^{ia^{2/m}z}-y)}}
\eqn\cxlviii $$}

\def\eqcxlviiia{
$$
x=1-a^{2/m}\up
\eqn\cxlviiia
$$}

\def\eqcxlviiib{
$$
y=y_c+a^{2/n}\um
\eqn\cxlviiib
$$}

\def\eqcil{
$$
\kappa\pt\uo (l)=-{1\over 2\pi i}
\int_{-i\infty}^{+i\infty}dz{e^{lz}\over\sz}
\eqn\cil
$$}

\def\eqcl{
$$
\uo (z)\equiv\kappa{\cal L}\lbrack\uo (l)\rbrack
\eqn\cl
$$}

\def\eqcli{
$$
\pt\uo (z)=-{1\over\sz}
\eqn\cli
$$}

\def\eqcliii{
$$
\pt\uo (l)=-{1\over\kappa}{e^{-l\up}\over\sqrt{\pi l}}
\eqn\cliii
$$}

\def\eqclv{
$$\eqalign{
&(x-y)\pxy\L =-\h (\px\L -\py\L )\cr
&(x-y)\pxy C =\h (\px C -\py C )\cr}
\eqn\clv
$$}

\def\eqclvi{
$$\eqalign{
&\px\L =\h (x-y)\px C\cr
&\py\L =\h (y-x)\py C\cr}
\eqn\clvi
$$}

\def\eqclva{
$$
\ppm t\sim -{1\over 2(1-y_c)}(a^{2/n}\pp t+a^{2/m}\pms t)
\eqn\clva
$$}

\def\eqclvii{
$$
\ppm t=0\; ,\qquad\ppm c=0
\eqn\clvii
$$}

\def\eqclviii{
$$
\pp c=\pp t\; ,\qquad\pms c=-\pms t
\eqn\clviii
$$}

\def\eqclix{
$$\eqalign{
&t=-\h\sum_{p\geq 0}(\tp _p\up^p +\tm_p\um^p)\cr
&c=-\h\sum_{p\geq 0}(\tp _p\up^p -\tm_p\um^p)=0\cr}
\eqn\clix
$$}

\def\eqclx{
$$
t=-\sum_{p\geq 0}\tp _p\up^p\; ,\qquad t=-\sum_{p\geq 0}\tm_p\um^p
\eqn\clx
$$}

\def\eqclxi{
$$
d^p_\pm t=2\p^p_\pm t
\eqn\clxi
$$}

\def\eqclxii{
$$
{\tilde t}^\pm _p=-{1\over p!}d^p_\pm t|_{u_{\pm}=0}
\eqn\clxii
$$}

\def\eqclxiii{
$$
d_+t|_0=\dots =d_+^{m-1}t|_0=0
\eqn\clxiii
$$}

\def\eqclxiv{
$$
{\p\up\over\p\tp_k}=\up^k\pt\up\; ,\qquad{\p\up\over\p\tm_k}=0
\eqn\clxiv
$$}

\def\eqclxiva{
$$
(x-y)\pxy\n =\h (\px\n -\py\n)
\eqn\clxiva
$$}

\def\eqclxivb{
$$
\ppm\tau =0
\eqn\clxivb
$$}

\def\eqclxv{
$$
\tau ={{\tilde\tau}_0\over 2}+\h\sum_{p > 0}(\Tp _p\up^p
+\Tm_p\um^p)
\eqn\clxv
$$}

\def\eqclxvia{
$$
u_2(p)\sim \h (1-y_c)^{3/2}a^{-2-3/m}{\n\pp\n\pt\up\over\tz}
\eqn\clxvia
$$}

\def\eqclxvii{
$$
\n =a^{2+1/m}{\sqrt{2}\tau\over 1-y_c}
\eqn\clxvii
$$}

\def\eqclxviii{
$$
\u_2 (z)={\tau\pp\tau\pt\up\over\tz}
\eqn\clxviii
$$}

\def\eqclxix{
$$
\pt\bigl(\u (z)-{\tau\pp\tau\pt\up\over\tz}\bigr) =-{1\over\sz}
\eqn\clxix
$$}

\def\eqclxx{
$$
\pt\bigl(\v (z)+{\tau\over\sz}\bigr) =2{\pp\tau\pt\up\over\sz}
\eqn\clxx
$$}

\def\eqclxxi{
$$
\v (z)\equiv a^{-2}\sqrt{1-y_c}{\v (p)\over \sqrt{2}}
\eqn\clxxi
$$}

\def\eqclxxii{
$$
\pt\bigl(\u
(l)-{2\over\kappa}\sqrt{l\over\pi}e^{-l\up}\tau\pp\tau\pt\up\bigr)
=-{e^{-l\up}\over\kappa\sqrt{\pi l}}
\eqn\clxxii
$$}

\def\eqclxxiii{
$$
\pt\bigl(\v (l)+{1\over\kappa\sqrt{\pi l}}\tau e^{-l\up}\bigr)
=2{e^{-l\up}\over\kappa\sqrt{\pi l}}\pp\tau\pt\up
\eqn\clxxiii
$$}

\def\eqclxxiiia{
$$
{d\over dg_k}\langle\dots\rangle
=-{N\over\L}\langle\sum_{i=1}^N\l_i^n\dots\rangle
\eqn\clxxiiia
$$}

\def\eqclxxiv{
$$
\langle\u (l)\dots\rangle =-{\L\over N}{d\over
dg_k}\langle\dots\rangle\equiv{\cal J}^B(l)\langle\dots\rangle
\eqn\clxxiv
$$}

\def\eqclxxv{
$$
{1\over\pi}\int_y^x{dp\;
p^k\over\sqrt{(x-p)(p-y)}}\sim{1\over\sqrt{x-y}}{x^{k+\h}\over\sqrt{
\pi k}}\; ,\qquad |x|>|y|
\eqn\clxxv
$$}

\def\eqclxxvi{
$$
{\p x\over\p g_k}\sim -{\sqrt{x-y}\over
4}\sqrt{k\over\pi}{x^{k-\h}\over\px\L}\; ,\qquad{\p y\over\p g_k}\sim
0\; ,\qquad |x|>|y|
\eqn\clxxvi
$$}

\def\eqclxxvii{
$$
{\cal J}^B(l)=-{\L\over N}{\p\up\over\p g_k}{\delta\over\delta\up}
\sim-\kappa\sqrt{l\over\pi}e^{-l\up}\pt\up{\delta\over\delta\up}
\eqn\clxxvii
$$}

\def\eqclxxviii{
$$
{\cal J}^B(z)=-{1\over 2}{\pt\up\over\tz}{\delta\over\delta\up}
\eqn\clxxviii
$$}

\def\eqclxxviiia{
$$
{\cal J}^F(l)=-a^{-1/m}{\L\over
N}{\partial\tau\over\p\xi_{k+\h}}{\delta\over\delta\tau}\sim-{\kappa
\over 4}{e^{-l\up}\over\sqrt{\pi l}}{\delta\over\delta\tau}
\eqn\clxxviiia
$$}

\def\eqclxxviiib{
$$
{\cal J}^F(z)={1\over 4}{1\over\sz}{\delta\over\delta\tau}
\eqn\clxxviiib
$$}

\def\eqclxxviiic{
$$
\k\pt^2F=-\up +2\pt (\tau\pp\tau\pt\up -\tau\pms\tau\pt\um )
\eqn\clxxviiic
$$}

\chapter{INTRODUCTION}

A description of the coupling of $\;N=1\;$ superconformal field
theories \REF\fqs{\rfqs}[\fqs] to two-dimensional supergravity
in terms of discrete triangulations of super-Riemann surfaces
is an interesting problem still unsolved. The continuum
analysis carried out in
\REF\pz{\rpz}[\pz]\REF\dhk{\rdhk}[\dhk] has not yet
found a complete discrete counterpart. The use of the
Kadomtsev-Petviashvilii (KP) hierarchy initiated by Douglas
\REF\douglas{\rdouglas}[\douglas] in the analysis of the double
scaling limit of the purely bosonic theories
\REF\doubles{\rdoubles}[\doubles] was not fully succesfully
pursued for the unitary
superconformal $(m,m+2)$ models in \REF\difdk{\rdifdk}[\difdk]
in terms of the Manin-Radul \REF\mr{\rmr}[\mr] supersymmetric
extension of the KP-equations. The root of the problem was the
incompatibility between the string equation and the fermionic
flows. In \REF\sloops{\rsloops}[\sloops] a proposal was made to
describe the coupling between $(2,4m)$ superconformal models and
world-sheet supergravity. The model is based on a set of superloop
equations which are motivated by analogy with the bosonic case
\REF\kazakov{\rkazakov}[\kazakov]. The planar solution to this
model allowed the construction of bosonic and fermionic
operators whose dressed gravitational dimensions are in
agreement with the continuum limit result for the
$(2,4m)$-models [\pz,\dhk]. The planar solution was constructed
for simplicity starting with even bosonic potentials. This as
expected generated a doubling of degrees of freedom in the
Neveu-Schwarz (NS) and Ramond (R) sectors of the theory. It was
also possible to obtain the correlation functions of an
arbitrary number of planar superloops.

The guiding principle in [\sloops] was a set of
super-Virasoro constraints satisfied by the partition
function, which code the superloop equations. Since the
Virasoro constraints \REF\dvv{\rdvv}[\dvv]\REF\fkn{\rfkn}[\fkn]
played a prominent role in the Witten-Kontsevich theory (the
intersection theory of certain line bundles on the moduli space
${\cal M}_{g,n}$ of genus g-surfaces with n distinguished
points) \REF\witkon{\rwitkon}[\witkon], it is reasonable to
expect that a similar set of super-Virasoro constraints should
capture some important features of the supermoduli space of
super-Riemann surfaces. The model obtained in this way is an
``eigenvalue'' model. It is formulated in terms of a collection
of $N$ even and $N$ odd eigenvalues $(\lambda_i,\theta_i)$. We
have still not succeded in finding a description of the model in
terms of generalized matrices, whose large$-N$ expansion may
provide a geometrical interpretation in terms of triangulated
super-surfaces. The aim of this paper is to continue the
analysis of this eigenvalue model. In section two we
briefly formulate the model and its loop equations, in section
three we solve it in the planar limit for general potentials. We
identify the scaling limit, the
 critical points and see how the doubling of degrees of
freedom is lifted for general potentials in section four. In section
five we obtain the continuum limit for bosonic and fermionic loops.
We
prove, that in the continuum limit they become respectively a
$Z_2$-twisted scalar and a Weyl-Majorana fermion in the Ramond
sector. Thus the super-Virasoro constraints in the double scaling
limit are described by the super-energy-momentum tensor of a
${\widehat c}=1$ superconformal field theory. In section six
we obtain the solution to the continuum loop equations in genus
zero, one and partially in genus two. For genus zero this provides a
good
verification of previous arguments. We find that the
terms independent of
fermionic couplings in the free energy coincides with the
result of the bosonic model. This result  is in slight conflict with
the genus one calculation of \REF\berkleb{M. Bershadsky and I.
Klebanov, {\it Nucl. Phys. }{\bf B360} (1990) 635}[\berkleb].
However, there was no clear way of normalizing the supermoduli
integration in [\berkleb], and therefore this may be the origin
of the discrepancy. Motivated by the agreement up to genus 	two
of the purely bosonic partition function of our model with
the one appearing in the Kazakov multicritical
points, we present a heuristic proof
that this equivalence works to all orders of string pertubation
theory.  On the basis
of our computations plus the
heuristic proof, we conjecture
that the KdV hierarchy describes the even
flows of our model when all fermionic couplings
are set to zero.  This
is indeed the hallmark for a supersymmetric extension of the
KdV hierarchy, although at present we have not been able
to identify it fully.
In section seven
we briefly review the difficulties we find in identifying our
bosonic and fermionic flows with known supersymmetric
extensions of the KdV- or KP-hierarchies
\REF\mulase{M. Mulase, ``A new super KP system
and a characterization of the Jacobians of arbitrary algebraic super
curves'', {\it J. Diff. Geom}, to appear}\REF\rabin{\rrabin}
\REF\skdv{\rskdv} [\mr,\mulase,\rabin,\skdv]. In [\sloops] the
Mulase-Rabin [\mulase] extension of KP seemed a promising candidate
to describe our model. Recently some work in this direction has
been done \REF\itoyama{H. Itoyama, {\it
Integrable Superhierarchy of Discretized 2d Supergravity}, Stony
Brook preprint, ITP-SB-92-21, June 1992, hepth-9206091.}[\itoyama]
\foot{In this paper the double scaling limit of the
super-Virasoro constraints is also obtained.}.
The difficulties we find
may be related to  the fact pointed out in [\rabin] that the
algebro-geometric solutions of the Mulase-Rabin or Manin-Radul
hierarchies do not describe the moduli space of super-Riemann
surfaces, but rather the moduli space of algebraic curves with genus
$g$ and a generic line bundle of degree $g-1$. In this sense, our
difficulties can be interpreted positively.  It may be disappointing
to find that the part of the free energy without fermion couplings
coincides with the results of one-matrix models.
Perhaps one should look at this result more positively, because it
indicates that our model does describe geometric objects
which extend fermionically Riemann surfaces, even though we
have not formulated the model in terms of triangulations.  The
kinds of geometries described by the Super-loop Equations
depend very much on the type of supersymmetric extension
of the KdV hierarchy emboddied by our model.
Section eight contains the conclusions and outlook.

\chapter{SUPERLOOP EQUATIONS}
We first review some properties of the purely bosonic one-matrix
models and see how they are generalized in the supersymmetric case.

The general one-matrix model partition function is
$$
Z=\int d^{N^2} \Phi exp[-{N\over \Lambda} tr V(\Phi)]\;\;,
$$
$$
V(\Phi)=\sum_{k \geq 0} g_k \Phi^k \qquad
\Lambda=e^{-\mu_B}\;\; .
\eqn\bi
$$
where $\Phi$ is a Hermitian $N\times N$ matrix, and $\mu_B$
is the bare cosmological constant. The starting point of
Kazakov's analysis of multicritical points
\REF\kazakov{\rkazakov}
[\kazakov] was a set of planar loop equations:

$$
\sum_{k\geq 1} k g_k {{\partial^{k-1}}\over{\partial l^{k-1}}}
w(l)=\int_0^l dl' w(l-l')w(l')\;\; ,
\eqn\bii
$$

$w(l)$ describes a loop of lenght $l$ bounding a surface with the
topology of a disk. Equation {\bii} is the planar limit of the
Schwinger-Dyson equations satisfied by the loop operator
$$
w(l)={\Lambda\over N} tr e^{l\Phi}= \sum_{n=0}^{\infty}
{l^n\over n!} w^{(n)}
\eqn\biii
$$
in the general one-matrix model {\bi}. Writing the partition
function in terms of the free energy $Z=e^{N^2 F}$,
$F=F_0+N^{-2}F_1+\dots$ the moments $w^{(n)}$ can be
represented as:
$$
w^{(0)}=\Lambda \qquad w^{(n)}=-\Lambda^2{\partial F\over
\partial g_n}\;\;,
\eqn\biv
$$
and the loop equations are equivalent to a set
of Virasoro constraints satisfied by {\bi} \REF\matsuo{Y.Matsuo,
unpublished}[\matsuo]. They are obtained by implementing
invariance of the partition function {\bi} under the change $\Phi
\rightarrow \Phi+\epsilon\Phi^{n+1},\;\;n\geq-1$ in \bi:
$$
L_n Z=0\quad n\geq -1\;\;,
\eqn\bv
$$

$$
L_n={\Lambda^2\over N^2}\sum_{k=0}^{n}{\partial^2\over\partial
g_{n-k}\partial g_k}+\sum_{k\geq 0}k g_k{\partial \over \partial
g_{k+n}}\;\;.
\eqn\bvi
$$
Using the Laplace transform of the loop operator {\biii},
$$
w(p)=\int_{0}^{\infty} e^{-pl}w(l) dl=\sum_{k=0}^{\infty}
{w^{(k)}\over p^{k+1}}\;\;,
\eqn\bvii
$$
and defining:
$$
\chi(p,q)=\sum_{k,l \geq 0}{\chi_{k,l}\over p^{k+1} q^{k+1}}
\qquad \chi_{k,l}=\Lambda^4{\partial^2 F \over \partial g_k
\partial g_l}
\eqn\bviii
$$
the loop equation, which is equivalent to
the Virasoro constraints {\bv} and {\bvi} is
$$
w(p)^2-V'(p)w(p)+{1\over N^2}\chi(p,p)={\rm Polynomial}(p)\;\;.
\eqn\bix
$$
It is easy to check that {\bix} in the limit
$N^2\rightarrow \infty$ becomes  {\bii}. The explicit form of the
polynomial on the right hand side is irrelevant. A
useful way of rewriting {\bix} is obtained after introducing an
infinite set of creation and anihilation operators:
$$
\alpha_{-n}=-{N\over \Lambda \sqrt{2}}n g_n,\quad n>0 \qquad
; \qquad \alpha_n =-{\Lambda \sqrt{2} \over N}{\partial \over
\partial g_n}, \quad n\geq 0\;\;,
\eqn\bx
$$
and a scalar field:
$$
\partial \varphi(p) =\sum_{n\in Z} \alpha_n p^{-n-1}
\eqn\bxi
$$
with energy-momentum tensor:
$$
T(p)={1\over 2}:\partial \varphi(p) \partial \varphi(p):
\eqn\xii
$$
Then, {\bix} becomes
$$
Z^{-1} T(p) Z={\rm Polynomial}(p)\;\;.
\eqn\bxiii
$$
Furthermore, if we write the partition function {\bi} in terms of
the eigenvalues of the $\Phi$-matrix $(\lambda_1,\lambda_2,\dots,
\lambda_N)$, but leaving the measure $\Delta^2(\lambda)$
undetermined :
$$
Z=\int \prod_{i} d\lambda_i \Delta^2(\lambda) e^{-{N\over
\Lambda} \sum_{i} V(\lambda_i)}\;\;,
\eqn\bxiv
$$
the constraints {\bv}, {\bvi} yield a differential equation
satisfied by $\Delta$
$$
\sum_{i}\lambda_i^{n+1}{\partial \Delta \over \partial
\lambda_i}= \Delta \sum_{i\neq j}
{\lambda_i^{n+1}\over{\lambda_i-\lambda_j}}\;\;,
\eqn\bxv
$$
whose solution up to a constant is the expected Van-der-Monde
determinant
$$
\Delta=\prod_{i<j} (\lambda_i-\lambda_j)\;\;.
\eqn\bxvi
$$
In the supersymmetric case we proceed by analogy [\sloops] with the
above arguments to obtain the corresponding loop equations.  The
loop operator depends on two variables $(l,\theta)$ where $l$ is
even and $\theta$ is odd. We can define the Laplace transform as
well
$$
w(p,\Pi)\equiv v(p)+\Pi u(p)=\int_0^{\infty}dl \int d\theta
e^{-pl-\Pi \theta} w(l,\theta)\;\;.
\eqn\bxvii
$$
In particular, the Laplace transform of the operator
$D={\partial \over \partial \Pi}+\Pi {\partial \over \partial
p}$ is ${\cal P}=\theta-z{\partial\over\partial\theta}$. Asumming
the loop $w(l,\theta)$ to behave well at $l=0,\infty$, we can
expand $w(p,\Pi)$ in inverse powers of $p$:
$$
v(p)=\sum_{k\geq 0}{v^{(k)}\over p^{k+1}}\qquad;\qquad
u(p)=\sum_{k\geq 0}{u^{(k)}\over p^{k+1}}\;\;,
\eqn\bxviii
$$
$v(p)$ and $u(p)$ are respectively the fermionic and bosonic
loops. To define the moments $v^{(k)}$, $u^{(k)}$ in terms of
the free energy $F={\ln Z\over N^2}$ we introduce bosonic and
fermionic oscillators:
$$
\alpha_p=-{\Lambda\over N}{\partial \over \partial
g_p}\qquad;\qquad \alpha_{-p}=-{N\over \Lambda}p g_p\quad,\quad
p=0,1,2,\dots
\eqn\bxix
$$
$$
b_{p+1/2}=-{\Lambda \over N}{\partial \over\partial \xi_{p+1/2}}
\qquad;\qquad b_{-p-1/2}=-{N\over \Lambda}\xi_{p+1/2}\quad,\quad
p=0,1,2,\dots
\eqn\bxx
$$
together with a free massless superfield:
$$
X(p,\Pi)=x(p)+\Pi \psi(p)
\eqn\bxxi
$$
$$
\partial X(p)=\sum_{n\in Z}\alpha_n p^{-n-1}\qquad
\psi(p)=\sum_{r\in Z+1/2} b_r p^{-r-1/2}\;\;,
\eqn\bxxii
$$
with the energy-momentum tensor:
$$
T(p,\Pi)\propto DX \partial X=\psi\partial_px+\Pi:(\partial_p x
\partial_p x+\partial_p\psi \psi):
\eqn\bxxiii
$$
The basic postulate in [\sloops] is to take the superloop equations
to be
$$
Z^{-1}T(p,\Pi)Z={\rm Polynomial}(p)\;\;.
\eqn\bxxiv
$$
In terms of $u(p)$, $v(p)$ these equations become:
$$
(u(p)-V'(p))^2+(v(p)-\xi(p))'(v(p)-\xi(p))+{\chi^{{BB}}(p,p)\over
N^2}+{\chi^{FF}(p,p)\over N^2}=Q_0\;\;, \eqn\bxxv
$$
$$
u(p)v(p)-V'(p)v(p)-\xi(p)u(p)+{\chi^{BF}(p,p)\over N^2}=
Q_1\;\;,
\eqn\bxxvi
$$
where:
$$
V(p)=\sum_{k\geq 0}g_k p^{k} \qquad \xi(p)=\sum_{k\geq 0}
\xi_{k+1/2} p^k
\eqn\bxxvii
$$

$$\eqalign{
&\chi^{BF}(p,q)=\sum_{k,l\geq 0}{\L^4 \over p^{k+1}
q^{l+1}}{\p^2 F \over \p \xi_{k+1/2} \p g_l}\cr
&\chi^{BB}(p,q)=\sum_{k,l\geq 0}{\L^4\over p^{k+1}
q^{l+1}}{\p^2 F\over
\p g_k\partial g_l}\cr
&\chi^{FF}(p,p)=\sum_{n\geq 1}\sum_{r=1/2}^{n-1/2}
{\L^4\over p^{n+2}}({n\over2}-r)
{\p^2 F\over\p \xi_r \p \xi_{n-r}}\;\;.\cr}
\eqn\bxxix
$$
The quantities $Q_0,Q_1$ are polynomials in $p$, and although
their explicit form can be computed, they will not be needed
throughout this paper. In terms of the original loop variables
$w(l,\theta)$ , the equations \bxxv , \bxxvi take a form
similar to \bii
$$
{\cal P}{\cal K} w(l,\theta)+2{\cal K} {\cal
P}w(l,\theta)=(w\circ {\cal P}w)(l,\theta)\;\;,
\eqn\bxxx
$$
with
$$
{\cal P}=\theta -l{\partial \over \partial \theta}
\eqn\bxxxi
$$
$$
{\cal K}\equiv \sum_{k\geq 1}(k g_k {\partial \over \partial
\theta}-\xi_{k-1/2}){\partial^{k-1}\over\partial l^{k-1}}\;\;,
\eqn\bxxxii
$$
where the convolution between two superfunctions $f_1(z,\theta)$,
$f_2(z,\theta)$ is defined according to:
$$
(f\circ g)(z,\theta)\equiv\int d\theta'\int_0^z f(z',\theta')
g(z-z',\theta-\theta') dz'\;\;,
\eqn\bxxxiii
$$
(see [\sloops] for more details). The previous arguments suggest,
in analogy with the one-matrix model, the introduction of a
``superpotential''
$$
V(\lambda,\theta)=\sum_{k\geq0}\sum_{i=1}^{N}(g_k\lambda_i^k
+\xi_{k+1/2}\theta_i\lambda_i^{k})\;\; .
\eqn\bxxxiv
$$
The moments $u^{(k)}$ and $v^{(k)}$ {\bxviii} can thus be
identified with derivatives of the free energy
$$
u^{(0)}=\L \qquad u^{(n)}=-\L^2{\p F\over \p g_n} \qquad
v^{(n)}=-\L^2{\p F \over \xi_{n+1/2}}\;\;.
\eqn\bxxxivbis
$$
Writing the partition function as:
$$
Z=\int \prod_{i=1}^N d\lambda_i d\theta_i \Delta(\lambda,\theta)
e^{-{N\over \Lambda} V(\lambda,\theta)}\;\;,
\eqn\bxxxv
$$
we can determine the explicit form of the measure
$\Delta(\lambda,\theta)$ by imposing the super-Virasoro
constraints {\bxxiv}. The explicit representation of the
super-Virasoro operators using $(2.19)$,
$(2.20)$ as differential operators is:
$$
G_{n-1/2}=\sum_{k=0}^\infty \xi_{k+1/2}{\partial \over
\partial g_{k+n}}+\sum_{k=0}^\infty kg_k{\partial\over\partial
\xi_{k+n-1/2}}+{\Lambda^2\over
N^2}\sum_{k=0}^{n-1}{\partial\over \partial
\xi_{k+1/2}}{\partial\over \partial g_{n-1-k}} \quad n\geq 0\;\;,
\eqn\bxxxvi
$$
$$
L_n={\Lambda^2\over 2N^2}\sum_{k=0}^n{\partial^2\over\partial
g_k \partial g_{n-k}}+\sum_{k=1}^\infty
kg_k{\partial\over\partial g_{n+k}}
$$
$$
+{\Lambda^2 \over
2N^2}\sum_{r=1/2}^{n-1/2}({n\over2}-r)
{\partial\over\partial\xi_{r}} {\partial\over\partial
\xi_{n-r}}+\sum_{r=1/2}^\infty({n\over2}+r)\xi_r{\partial\over
\partial\xi_{r+n}}\quad n\geq -1\;\;.
\eqn\bxxxvii
$$
Since $\{G_{n-1/2},G_{m-1/2}\}\propto L_{n+m-1}$, it suffices to
implement $G_{n-1/2} Z=0$. This leads to a set of equations:
$$
\sum_{i}\lambda_i^n(-{\partial\over\partial\theta_i}
+\theta_i{\partial\over\partial\lambda_i})\Delta=\Delta
\sum_{i\neq j} \theta_i{\lambda_i^n-\lambda_j^n\over
\lambda_i-\lambda_j}\;\; ,
\eqn\bxxxviii
$$
whose unique solution, up to a multiplicative constant is:
$$
\Delta(\lambda,\theta)=\prod_{i<j}(\lambda_i-\lambda_j
-\theta_i\theta_j)\;\;.
\eqn\bxxxix
$$
Hence the model we would like to solve in the large-$N$ limit is:
$$
Z=\int \prod_{i=1}^Nd\lambda_i d\theta_i \prod_{i<j}(\lambda_i
-\lambda_j-\theta_i\theta_j)e^{-{N\over\Lambda}V(\lambda,\theta)}
\;\;.
\eqn\bxl
$$
The loop operator can be explicitly written as:
$$
w(l,\theta)\equiv{\Lambda\over N}\sum_{i} e^{l\lambda_i+\theta
\theta_i}\;\;.
\eqn\bxli
$$
 From {\bxl} and {\bxli} one can derive {\bxxx}. The
simplifying assumption made in [\sloops] was $g_{2k+1}=0,\;\;
k\geq 0$; i.e. the bosonic part of the potential {\bxxxiv} was
taken to be even. In the next section we begin the analysis of
{\bxl} without this restriction.

\chapter{SOLUTION TO THE PLANAR LOOP EQUATIONS: GENERAL
POTENTIAL}
In this section we study the loop equation {\bxxv}, {\bxxvi} in
genus zero for an arbitrary bosonic part of the super-potential. The
simplifying assumption $g_{2k+1}=0$  made in [\sloops] generated a
doubling of degrees of freedom in the Neveu-Schwarz and Ramond
sectors of the theory that is not present in the continuum
super-Liouville theory  \REF\sliouv{\rsliouv}[\sliouv].
The planar loop equations follow from {\bxxv},{\bxxvi}:
$$
(u(p)-V'(p))^2+(v(p)-\xi(p))'(v(p)-\xi(p))=Q_0(p)\;\;,
\eqn\ci
$$
$$
(v(p)-\xi(p))(u(p)-V'(p))=Q_1(p)\;\;.
\eqn\cii
$$
Using the fact that $Q_1$ is fermionic, the solution to {\ci},
{\cii} is:
$$
u(p)-V'(p)=\sqrt{Q_0(p)}-{Q_1'(p)Q_1(p)\over2Q_0(p)^{3/2}}
\eqn\ciii
$$
$$
v(p)-\xi(p)={Q_1(p)\over\sqrt{Q_0(p)}}\;\;.
\eqn\civ
$$
As in the pure gravity case we look for the one-cut solution.
Since we make no assumptions concerning the parity of $V(p)$, the
one-cut solution takes the form:
$$
u(p)=u_0(p)+u_2(p)=V'(p)-M(p)\sqrt{\Delta}-{A(p)\over\D^{3/2}}
\eqn\cv
$$
$$
v(p)=\xi(p)-{N(p)\over\sqrt{\Delta}}
\eqn\cvi
$$
with $\D=(p-x)(p-y)$. The subindex in {\cv} indicates the order
in fermionic couplings. We can also introduce variables $R,S$:
$$
x=S+\sqrt{R}
$$
$$
y=S-\sqrt{R}\;\;.
\eqn\cvii
$$
For $V(p)=V(-p)$ the cut is symmetric and $S=0$. Since $u(p)\sim
O(1/p)$, $v(p)\sim O(1/p)$ as $|p|\rightarrow \infty$,
$M(p),N(p)$ are determined as functions of $V'(p)$ and $\xi(p)$
respectively directly from this requirement. After $M$ and $N$ are
determined, the form of $A(p)$ follows from demanding that the
left hand sides of {\ci} and {\cii} must be polynomials. To
write down the explicit form of $M$, $N$ and $A$ we note, that
any analytic function $f(p)$ can be written in the form:
$$
f(p)=f_0(\D)+\D'f_1(\D)\;\;,
$$
$$
\D'={d\D\over dp}=2p-x-y=2(p-S)\;\;.
\eqn\cviii
$$
We split $f(p)$ into two terms with opposite parity with
respect to the change $(p-S)\rightarrow (S-p)$. Hence:
$$\eqalign{
&M(p)=M^-(\D)+\D'M^+(\D),\cr
&N(p)=\D N^-(\D)+\D'N^+(\D),\cr
&A(p)=A^-(\D)+\D'A^+(\D)\;\;.\cr}
\eqn\cix
$$
In $N(p)$ we have used the fact that our solution should agree
with the result obtained in [\sloops], when $V(p)=V(-p)$ \foot{To
obtain the even case a factor of two must be taken into account as a
normalization of the $n_k^{+}$'s in $(3.10)$}. The expansions of $M$,
$N$ and $A$ in powers of $\D$ are given by:
$$
M^{\pm}(\D)=\sum_{k\geq 0} m_{k}^{\pm}\D^k, \qquad
N^{\pm}=\sum_{k\geq 0}n_{k}^{\pm}\D^k, \qquad
A^{\pm}(\D)=\sum_{k\geq 0}a_{k}^{\pm}\D^k\;\;.
\eqn\cx
$$
To determine $A$, we substitute these expressions in {\cv},
{\cvi} and require the left-hand side of {\ci} and {\cii} to be
polynomials in p. After some computations, we obtain that
$A^-(\D)$ and $A^+(\D)$ are completely given by $a_0^{-}$,
$a_0^{+}$. The results are
$$
A^-=-{2R\over ({m^{-}_0})^2-4R({m^{+}_0})^2}({m_0^{-}}n_0^{-}
n_0^{+}-4Rm_0^{+}n_1^{+}n_0^{+})\;\;,
\eqn\cxi
$$
$$
A^+=-{2R\over({m^{-}_0})^2-4R({m^{+}_0})^2}({m_0^{-}}n_1^{+}
n_0^{+}-m_0^{+}n_0^{-}n_0^{+})\;\;.
\eqn\cxii
$$
Once we determine $M(p)$ and $N(p)$ we will have the complete
solution to
the planar model. Notice that $N(p)$ will be linear in the fermionic
couplings. This implies that the non-vanishing planar
correlators contain at most two fermionic operators. This also
occurs in the even case $V(p)=V(-p)$, so
that this rather surprising phenomenon does not depend on the type
of potential chosen. We refer the reader to [\sloops] for further
discussions.

The solution to the planar model given by eqs. {\cv} and {\cvi} is
parametrized by $x$ and $y$. Since there is only one single physical
parameter $\Lambda$, we should be able to express both $x$ and $y$ as
functions of $\Lambda$.
Let $u_0(p)$ be the
purely bosonic part of the loop operator (no dependence on the
$\xi_{k+1/2}$ couplings). Then the one-cut solution is:
$$
u_0(p)=V'(p)-M(p)\sqrt{(p-x)(p-y)}\;\;.
\eqn\cxiii
$$
Since $V'(p)$ and $M(p)$ are polynomials in $p$, $u_0(p)$ admits the
following expansion
\eqcxiv
Dividing this expression by $\Delta^{1/2}$ gives
\eqcxv
which implies
\eqcxvi\eqcxvii
But according to {\bxviii}, $u_0(p)\sim 0(1/p)$. Thus we have the
following constraint on $x$ and $y$
\eqcxviia
Equations (3.17) and (3.18) can be used, in principle, to rewrite
the solution in terms of the single physical parameter $\Lambda$.
These equations have appeared previously in the matrix model
literature (see for instance \REF\kostov{I.Kostov, Preprint
Saclay 1991, SPhT/91-142}[\kostov] and references therein) in a
slightly different form. It is important to notice that the
derivation
is not based on the method of orthogonal polynomials
\REF\am{\ram}[\am] since this formalism is still missing in
the supersymmetric case.

Note that the piece proportional to $\h (x+y)$ in {\cxvii} vanishes
when {\cxviia} is enforced. Then one recovers the expression for
$\Lambda$ which is usually found in the literature. However, in order
to compute partial derivatives it is essential to use the complete
form {\cxvii}.

Expanding $M$ in powers of $p$
\eqcxviii
we find
\eqcxix
and
\eqcxx
In order to evaluate {\cxi}, {\cxii}, we also need explicit formulae
for
$m_0^\pm$. These can be obtained as follows. Differenciating {\cxvii}
and
comparing the result with {\cxx} gives
\eqcxxa
On the other hand,  by {\cx}
\eqcxxb
This gives $m_0^\pm$ in terms of the single function $\Lambda (x,y)$
\eqcxxc
We conclude our analysis of the purely bosonic part of the solution
by
obtaining a simple expression for the derivative of $u_0$ with
respect
to the cosmological constant $\Lambda$. Under a variation $(dx, dy)$
compatible with the constraint {\cxvi}, $du_0$ is given by
\eqcxxvi
Since the expression multied by $\Delta^{-1/2}$ is a polynomial, we
must have
\eqcxxvii
Therefore:
$$
{\p u_0(p)\over \p \L}={1\over \sqrt{(p-x)(p-y)}}
\eqn\cxxviii
$$

When we discuss the continuum limit, the following loop operator will
be relevant
\eqcxxix
Note that this also satisfies
\eqcxxx
We now turn to the fermionic part of the solution. It is convenient
to
rewrite {\cvi} in a slightly different form. Defining
\eqcxxxiv
eq. {\cvi} becomes
\eqcxxxv
where
\eqcxxxva
and
\eqcxxxvb
We can also find an expression analogous to {\cxxviii}.
Differentiating
{\cxxxv} with respect to $x$ gives
\eqcxxxvi
On the other hand, since by {\bxviii} $v(p)\sim 0(1/p)$,
\eqcxxxvii
This implies
\eqcxxxviii
and if we set $p=y$ we find $a=(x-y)b$.
Since
\eqcxl
comparison with {\cxxxvii} yields $b=\p_x n_0^+$.

We finally obtain
\eqcxlii
Interchanging $x$ and $y$ gives
\eqcxliii
In order to complete the planar solution, we need explicit formulae
for
the functions $n_0^-$  and $n_1^+$ appearing in {\cxi},{\cxii}.
First,
note that eq. {\cxxxviii} can be written
\eqcxliia
Setting $p=x$ and using {\cxxxiv} gives
\eqcxliib
This, together with
\eqcxliibb
(obtained from {\cxliii}) can be used to determine $n_0^-$ and
$n_1^+$
as functions of the single function $n_0^{+}$
\eqcxliic
As in the bosonic case, we may define $\v (p)\equiv v(p)-\xi(p)$,
which satisfies relations identical to {\cxlii} and {\cxliii}.
At this point we have determined the complete solution to the planar
model. In the next section we will consider the corresponding
continuum limit.

\chapter{THE SCALING LIMIT}
The definition of the double scaling limit with an arbitrary
potential
involves some subtle considerations. These can be best understood by
studying first the purely bosonic part of the theory. We take  as the
fundamental geometric `observable' the macroscopic loop, defined by
\eqcxlv
with $na^{2/m}=l$ fixed. Here $m$ is a positive integer related to
the
order of criticality. The continuum limit should be taken in such a
way that {\cxlv} makes sense. This will be our guiding principle.
Eq. (3.29) implies
\eqcxlvi
where the contour $C$ encloses the cut $(y,x)$. We may assume
$|y|<|x|\leq1$ without loss of generality. We will comment on the
symmetric
case $y=-x$ later. Taking $C$ as the unit circle around the origin
with $p=e^{i\theta}$,
\eqcxlvii
We set $\Lambda_c=1$, with $\Lambda=1-a^2 t$, where $t$ is the
renormalized cosmological constant. $x$ and $y$ will approach the
critical values $x_c$ and $y_c$, and $N$ will be related to the
renormalized string coupling constant $\kappa$ by
\eqcxlviib
(The factor $(x_c-y_c)^\h$ is introduced for later convenience).
Changing to the new variable $z=a^{-2/m}\theta$, eq. {\cxlvi} becomes
\eqcxlviii
This will vanish unless the integral itself is of order $a^{-1/m}$,
i.e., if $x$ approaches $1$ as $a^{2/m}$,
\eqcxlviiia
Since $x$ and $y$ are not independent variables,
$y$ will approach its critical value $y_c$ ($|y_c|<1$) at the same
time
\eqcxlviiib
Here $n$ is another positive integer, which may be different from
$m$.
The integral in {\cxlviii} is dominated by the region $p\sim x_c=1$,
and
the contour can be deformed into a straight line
\eqcil
Here we recognize the definition of the inverse Laplace transform.
Thus,
if we define
\eqcl
we have
\eqcli
It is interesting to note that this result can also be obtained from
{\cxxviii} by  a scaling $p=1+a^{2/m}z$ in $\hat u_0 (p)$, together
with (4.6) and (4.7).

There are two
different ways of viewing $\uo (z)$: as the Laplace transform of
the macroscopic loop $\uo (l)$, or as the continuum limit of the loop
operator $\uo (p)$.

 From {\cil} one gets the usual expression for the macroscopic loop
\eqcliii
Note that $\uo (l)$ is independent of the scaling variable $u_-$.
This result holds as long as $|y_c|<|x_c|$, independently of the
values of $m$ and $n$. It is easy to see that, for $|y_c|=|x_c|$, the
dominant endpoint will be the one with the highest order of
criticality.
For a symmetric potential both endpoints contribute, and we get the
familiar phenomenom of `doubling'.
As we shall see below, the situation is not so simple when one
considers
the fermionic contributions to the loop operators.

In order to complete our description of the continuum limit for the
purely bosonic part of the theory, we must consider the scaling of
equations {\cxvi} and {\cxvii}. Comparing  first derivatives of
$\Lambda(x,y)$ and $C(x,y)$, we find that they are not independent.
Instead,
\eqclvi
Similarly, we have the following identities
\eqclv
Scaling {\clv} according to {\cxlviiia} and {\cxlviiib} gives
\eqclva
with an identical expression for $C$. For $a\to 0$ the RHS vanishes,
and
we find
\eqclvii
where we have defined $\p_\pm\equiv {\p\over \p u_\pm}$ and
$c\equiv -{1\over 2}a^{-2}(1-y_c)C$. Moreover, {\clvi} implies
\eqclviii
The general solution to {\clvii} and {\clviii} can be written
\eqclix
where $\tilde t^\pm_p$ are the renormalized couplings. Eqs. {\clix}
are
the continuum version of {\cxvi} and {\cxvii}. Adding and subtracting
the equations in {\clix} give rise to two decoupled string equations
\eqclx
Comparing {\clix} and {\clx} we see that
\eqclxi\eqclxii
The total derivatives in {\clxii} are computed for variations
$du_\pm$
consistent with the constraint {\clix}. The relative factor of $2$
between total and partial derivatives is important. It means that we
can not set  $\p_- t=0$  consistently.

The $m$-multicritical point (at  $x$)  is defined by
\eqclxiii
One can impose similar constraints at $y$, with a different index
$n$.
But, as mentioned above, for $|y|<|x|$ the continuum limit is
controled
by $m$ (the converse is of course true for $|y|>|x|$).

Eq. {\clx} can be used to compute the derivatives of $u_+$ with
respect
to the renormalized couplings
\eqclxiv
These expressions will be useful in connection with the definition of
the free energy.

We now turn our attention to the fermionic contributions. The
following identity can be derived for $n_0^+$
\eqclxiva
and, if $\tau$ is the scaling function corresponding to  $n_0^+$, we
have
\eqclxivb
which implies
\eqclxv
This defines the renormalized fermionic couplings $\tilde\tau^\pm_n$.
Using eqs. (3.5), {\cxi}, {\cxii}, {\cxxc}, {\cxliic} we find the
following expression for $u_2$
\eqclxvia
The fact that $u_0$ and $u_2$ must scale in the same way fixes the
scaling for $n_0^+$
\eqclxvii
Then
\eqclxviii
and we get for the bosonic loop
\eqclxix
In order to obtain the fermionic loop we first note that, in the
continuum limit, the RHS of {\cxliii} vanishes. This turns the
partial
derivative in {\cxlii} into a total derivative $d_x$, and from
{\cxlii} we can write
\eqclxx
where we have defined
\eqclxxi
The different powers of $a$ in the definitions of the continuum
bosonic and fermionic loops are a consequence of their different
scaling dimension. The macroscopic loops are obtained by the
inverse Laplace transform. The result is
\eqclxxii\eqclxxiii
Note that the fermionic contributions to the loops are not
independent of $u_-$. In other words, even though the loops are
dominated by the contribution from $p\sim x_c$, their expectation
values depend on all the couplings, even those defining the
behaviour of the potentials at $y$. This peculiarity of the
supersymmetric theory implies that one has to be careful when
trying to obtain the double scaling limit of the SuperVirasoro
constraints. This will be our main concern in section 5.

Multiloop correlators are obtained by acting on $\u (l)$ and $\v (l)$
with the appropriate `loop insertion operators'.
We consider first the bosonic case. Since
\eqclxxiiia
we have
\eqclxxiv
We need to know the derivatives of $x$ and $y$ with respect to $g_k$.
These are obtained by differenciation of {\cxvi} and {\cxvii}. Since
we
are interested in the limit $k\to\infty$, we need the asymptotic
behavior
of integrals like {\cxlvi}. Converting  the integral around the cut
into
an integral along the real axis, and using Laplace method for
integrals
dominated by endpoint contributions, we find
\eqclxxv
and
\eqclxxvi
These formulae can be used for an alternative derivation
of the macroscopic loop. Scaling this expression gives the
`macroscopic loop insertion operator'
\eqclxxvii
It is important to note that this expression assumes that all the
dependence on the couplings is given implicitly through $u_+$ and
$u_-$. But one can always rewrite the expressions so that this is
actually true.
The Laplace transform of {\clxxvii} is
\eqclxxviii
This operator inserts a loop $\u(z)$.

The fermionic insertion operators are obtained along the same lines.
The result is
\eqclxxviiia
and
\eqclxxviiib
It is easy to see that the macroscopic loops can be obtained by
acting
with ${\cal J}^B(l)$ and ${\cal J}^F(l)$ on the following free
energy
\eqclxxviiic

\chapter{CONTINUUM SUPER-VIRASORO CONSTRAINTS}
We will now see that, in the continuum limit, the bosonic and
fermionic loop operators become a $Z_2$-twisted scalar
bosonic field and a Weyl-Majorana fermion in the Ramond sector.
This is the supersymmetric generalization of the well known result
for
the bosonic matrix model. However, proving this statement turns out
to
be rather subtle in our case. The reason is that, as mentionned in
the last section, the correlators depend on all the couplings,
not just those describing the continuum limit at the dominant
endpoint.
It is not a priori clear  that one can use  a single
$Z_2$-twisted  bosonic field and a single Weyl-Majorana fermion.
However, this turns out to be true.

First we will show that $\uo (z)$ can be identified with
a free bosonic scalar field $\varphi(z)$ in two dimensions with
antiperiodic boundary conditions
$$
\p\varphi(z)=\sum_{n\in Z} \alpha_{n+1/2}z^{-n-3/2} \quad ,\quad
\varphi(e^{2\pi i}z)=-\varphi(z)\;\; .
\eqn\di
$$
The Laplace transform of the bosonic loop
can be decomposed into two pieces
$$
\u (z)\equiv t(z)+u(z)
\eqn\dii
$$
where
$$
u(z)=\k\sum_{k\geq 0}{\langle \sigma_k^{+}\rangle \over z^{k+3/2}}
\qquad,\qquad
\langle \sigma_k^{+}\rangle ={\p F\over \p t_k^{+}}
\eqn\diii
$$
and $t(z)$ is the non-universal part. $t(z)$ admits an
 expansion in powers
$z^{n-1/2}$, for $n\geq 0$.

Eq. {\diii} is in fact a definition of
the scaling operators $\sigma_k^{+}$. Since all
the contributions to
the loop come from $p\sim x_c=1$, it makes
sense to expand in terms of
operators associated with the $x$-endpoint.
 Comparing {\clxix} with {\diii} implies
 $$
\pt\left( u(z)-{\tau\pp\tau\pt\up\over\tz}\right) =
-{1\over\sz}+{1\over\sqrt
z}
\eqn\diiia
$$
With the  Laurent expansion of {\diiia} for $z\rightarrow
\infty$:
$$
{\p u(z)\over \p t}=-
\sum_{k\geq 0}{(-1)^{k+1}\Gamma(k+{3\over 2})\over k!
\Gamma({1\over 2})}
\left( {u^{k+1}_{+}\over k+1}+2\p_t(u^k_{+} \tau \pp \tau\pt\up )
\right)z^{-k-3/2} \;\; .
$$
the scaling operators $\langle \sigma_k^{+}\rangle $ are
given by:
$$
\p_t \langle {\sigma}_k^{+}\rangle ={1\over \k}{(-1)^{k} \Gamma
 (k+{3\over 2})\over
k!\Gamma({1\over 2})}\left({\up^{k+1}\over k+1}+2\p_t
(\up^k\tau\pp\tau\pt\up ) \right) \;\; ,
\eqn\dv
$$
This result should also be obtained from the
free energy (4.42) by
differentiation with respect to $t_k^+$. Then {\dv}
implies the following
 relation between $t_k$ and ${\widetilde t}_k$
$$
{\tilde t}_k^{+}={(-1)^{k+1}\Gamma(k+{3\over 2}) \over
k!\Gamma({1\over 2})} t_k^{+} \;\; .
\eqn\dviii
$$
We now turn to the computation of $t(z)$. Since $(z+u_+)^{-3/2}$
can not contribute to the
non-universal part, it is obvious that $t(z)$ is a
function of $u_+$, independent of the fermionic couplings. Then,
subtracting {\diiia} from {\clxix} gives

$$
\pt t(z)={\p t(z)\over\p{\tilde t}^{+}_0}=-{1\over\sqrt z}
\eqn\dviiia
$$
where we have used
$$
\pt\up ={\p\up\over\p{\tilde t}^{+}_0}
\eqn\dviiib
$$
Thus $t(z)$ satisfies
$$
t(z)=-{{\tilde t}^{+}_0\over\sqrt z}+{\rm (indep. of}\; \up)
\eqn\dviiic
$$
On the other hand, according to (3.28),  $\uo (z)$ has the following
form
$$
\uo (z)=-M(z)\sqrt\D\equiv\sum_{n\geq 0}\uo ^n\D^{n+\h}
\eqn\dviiie
$$
where $M(z)$ is defined as the scaling limit of $M(p)$, and
$\Delta=z+u_+$. For $u_+=0$, $\uo (z)$ contains only positive half
integral powers of $z$.
This implies
$$
t(z)=\uo (z)|_{\up =0}-{{\tilde t}^{+}_0\over\sqrt z}
\eqn\dviiid
$$
Differentiating (5.10) with respect to $t$ and taking
into account that $\partial_t\uo=-1/\sqrt{\Delta}$, and using (5.7)
yields:
$$
\uo ^0=2d_+t\; ,\qquad d_+\uo^{n-1}=-(n+\h)\uo^n
\eqn\dviiif
$$
and we find the following expression for $\uo (z)$
$$
\uo (z)=\sum_{n\geq 0}(-1)^{n+1}{\Gamma (\h )\over\Gamma(n+{3\over
2})}(d_{+}^{n+1}t)(\zp )^{n+\h}
\eqn\dviiig
$$

Setting $u_+=0$ gives
$$
t(z)=-{{\tilde t}^{+}_0\over\sqrt z}+\sum_{n\geq 0}(-1)^{n+1}
{\Gamma (\h )\over\Gamma(n+{3\over
2})}(d_{+}^{n+1}t)|_{\up =0}z^{n+\h}
\eqn\dviiih
$$
and using {\clxii}, {\dviii} we finally get
$$
t(z)=\sum_{k\geq 0}(k+{1\over 2})t_k^{+} z^{k-1/2}\;\; .
\eqn\dix
$$
In total
$$
\u (z)=\sum_{k\geq 0}(k+{1\over 2})t_k^{+}z^{k-1/2}+\k \sum_{k\geq 0}
{\langle \sigma_k^{+}\rangle \over z^{k+3/2}}\;\; .
\eqn\dx
$$
The relations between the coupling constants of the model and
the modes of the bosonic field are
$$
\alpha_{n+1/2}=\kappa {\p \over \p t_n^{+}} \qquad ,\qquad
\alpha_{-n-1/2}=(n+{1\over 2}){t_n^{+}\over \kappa}
\qquad\qquad n\geq 0
\eqn\dxi
$$
so that we can write
$$
\u (z)=\kappa Z^{-1}\p\varphi (z)Z
\eqn\dxii
$$
We proceed analogously with the fermionic loop. First, decompose into
universal and non-universal parts
 $$
\v (z)\equiv\eta (z)+v(z)
\eqn\dxvi
$$
with

$$
v (z)=\k\sum_{n \geq 0}{1\over z^{n+1/2}}\langle
\nu_n^{+}\rangle\; ,\qquad\langle\nu_n^{+}\rangle =
{\p F\over\p\tau_n^{+}}
\eqn\dxvii
$$
${\widehat v(z)}$ will be identified with a Weyl-Majorana fermion
in the Ramond sector.
The Laurent expansion of {\clxx} determines the correlator of the
scaling
operators
$$
\pt\left( (-1)^{k+1}{k!\Gamma (\h )\over\Gamma (k+\h
)}\langle\nu_k^{+}\rangle +{1\over\k}\up^k\tau\right)
={2\over\k}\up^k\pp\tau\pt\up
\eqn\dxviia
$$
and the relation between ${\widetilde \tau_k}$ and $\tau_k$ is

$$
{\widetilde \tau_k^{+}}={(-1)^{k+1}\over k!}{\Gamma(k+{1\over 2})
\over \Gamma({1\over 2})} \tau_k^{+}\;\; .
\eqn\dxviii
$$

The computation of $\eta(z)$ follows closely the method used with
$t(z)$,
and we simply quote the result
 $$
\eta(z)={\tau_0\over 2\sqrt{z}}+\sum_{k\geq 0}\tau_{k+1}^{+}
z^{k+1/2}\;\; .
\eqn\dxix
$$
This makes it posssible to identify $\v (z)$ as a Weyl-Majorana
fermion in the Ramond sector. We can write
 $$
\v (z)=\kappa Z^{-1}\psi (z)Z
\eqn\dxx
$$

In terms of the mode expansions (5.20) and (5.23),
$$\eqalign{
&\psi(z)=\sum_{n\in Z}\psi_n z^{-n-1/2}\;\; ,\cr
&\psi_n=\kappa {\p \over \p \tau_n^{+}} \qquad,\qquad
\psi_{-n}={\tau_n^{+}\over \kappa} \qquad; \qquad n>0\cr}
\eqn\dxxi
$$
while for the zero mode we have,
$$
\psi_0={\tau_0\over 2\kappa }+\kappa {\p \over \p \tau_0}\;\; ,
\eqn\dxxii
$$
guaranteeing $\psi_0^2=\h$.

We have thus succeded in writing $\u(z)$ and $\v (z)$ entirely
in terms of the couplings associated with a single endpoint. One can
easily
check that the loop insertion operators constructed in the last
section can
be written as

$$
{\cal J}^B(z)=\sum_{n\geq 0}z^{-n-{3\over 2}}{\p\over\p t_n^{+}}
\eqn\dxxiia
$$
and
$$
{\cal J}^F(z)=\sum_{n\geq 0}z^{-n-\h}{\p\over\p\tau_n^{+}}
\eqn\dxxiia
$$
This means that the couplings $t_n^-$ and $\tau_n^-$ merely
parametrize the
loop correlators, but do not contribute to the `dynamics'.

The super-Virasoro constraints in the continuum are therefore
described by the super-energy momentum tensor of a single ${\widehat
c}=1$ superconformal field theory,
$$
T_F(z)={1\over 2}\p\varphi(z)\psi(z)\;\; ,
\eqn\dxxiii
$$
$$
T_B(z)={1\over 2}:\p \varphi (z)\p \varphi (z):+{1\over 2}:\p
\psi(z) \psi(z):+{1\over 8z^2}\;\; .
\eqn\dxxiv
$$
With the mode expansion
$$
T_F(z)=\h\sum_{n \in Z}{G_{n+1/2}\over z^{n+2}}
\quad , \quad T_B(z)=\sum_{n\in Z}{L_{n}\over z^{n+2}}\;\; ,
\eqn\dxxv
$$
we obtain, in terms of the coupling constants:
$$
G_{n+1/2}={t_0^{+}\tau_0\over 4\k}\delta_{n,-1}+
\sum_{k\geq 0}(k+{1\over
2})t_k^{+}{\p \over \p \tau_{n+k+1}^{+}}+ {\tau_0\over 2}{
\p  \over \p
t_n^{+}}+ \sum_{k\geq 0}\tau_{k+1}^{+}{\p \over \p t_{n+k+1}^{+}}
$$
$$
+\k\sum _{k=0}^{n}{\p^2\over \p t_k^{+} \p \tau_{n-k}^{+}}\;\; ,
\eqn\dxxvii
$$

$$
L_n={{t_0^{+}}^2-2\tau_0\tau_1^{+}\over 8\k}\delta_{n,-1}+
\sum_{k\geq 0}
(k+{1\over 2})t_k^{+}{\p \over \p t_{n+k}^{+}}+{\k\over 2}\sum
_{k=1}^{n}{\p^2\over \p t_{k-1}^{+} \p t_{n-k}^{+}}
$$
$$
+{n\over 4}\tau_0{\p\over\p\tau_n^{+}}+\sum_{k\geq
0}({n\over 2}+k+1)\tau_{k+1}^{+}{\p  \over \p \tau_{n+k+1}^{+}}
-{\k\over
2}\sum_0^n(k+\h){\p\over\p\tau_k^{+}}{\p\over\p\tau_{n-k}^{+}}
+{1\over
8}\delta_{n,0}
\eqn\dxxviii
$$
They satisfy the conmutation relations
$$
\{G_m,G_n\}=2L_{n+m}+{1\over 2}(m^2-{1\over
4})\delta_{m+n,0}
\eqn\dxxvii
$$
In fact, the non-universal term ${1\over 8}\delta_{n,0}$ is fixed by
these relations.

The basic postulate of [\sloops] was to take the
discrete superloop equations equivalent to:
$$
Z^{-1}T(p,\Pi) Z={\rm Polynomial} (p,\Pi)
\eqn\dxxviii
$$
After the double scaling limit we obtain that the superloop
equations in the continuum are equivalent to:
$$
Z^{-1}T(z)Z={\rm Polynomial}(z)
\eqn\dxxix
$$
with $T(z,\theta ) =T_F(z)+\theta T_B(z)$ given by {\dxxiii} and
{\dxxiv}
This proves one of the main results of this paper. The continuum
limit of our superloop equations are described by a $Z_2$-twisted
massless scalar field, and a Weyl-Majorana fermion in the Ramond
sector.

We would like to remark
that the superVirasoro operators act on $Z$ instead of $\sqrt Z$, as
happens for cases with doubling. This fact was conjectured in
[\dvv], but we
do not know of any previous complete proof of this statement.

We finish this section with a comment about the dimensions of scaling
operators. With a similar calculation as in [\sloops] we obtain for
$\langle \sigma_k^{+}\rangle,\;\; d_k=k/m$ and for $\langle \nu
_k^{+}
\rangle \;\; d_k=k/m-1/2m$. These are the gravitational scaling
dimensions of operators in the NS- resp. R- sector of $(2,4m)$ $N=1$
superconformal minimal models coupled to
two dimensional gravity. Since
we have considered generic potentials
there appears no doubling of
operators as it was in [\sloops].

\chapter{SUPERLOOP EQUATIONS IN THE CONTINUUM}
In the following we solve the continuum loop equations
determined by {\dxxix}. Here the formulae
get more transparent than in the discrete theory. It is interesting
to consider also the planar case and see how the results for the
loop operators and the string equation found in section four
appear in a simple way. For higher genera, i.e. for the
torus, the double-torus, $\dots$  we obtain a sytematic expansion
determining all correlators beyond the planar limit.  In the
arguments of this section we are going to take for
simplicity $\tau (u_+,u_-)$ as a function only of
$u_+$.  In other words, we set $\tau_k^{-}=0$ for all $k$.  We
first show how the loop equations are solved up to genus
two, and then present a general heuristic argument showing that
the purely bosonic part of our model is equivalent to
the KdV-hierarchy.

The two superloop equations in the double scaling limit, which
are equivalent to the continuum super-Virasoro constraints
are:
$$
\u(z)\v(z)+\k\chi^{BF}(z)={\rm
Polynomial}(z) ,
\eqn\ei
$$
$$
\u(z)^2-\v(z) \p \v(z)
+\k(\chi^{BB}(z)+\chi^{FF}(z)+
{1\over 4z^2})={\rm Polynomial}(z),
\eqn\eii
$$
where the two-loop operators are defined by:
$$\eqalign{
&\chi^{BF}(z)= \sum_{k,l\geq 0}{1\over
z^{k+l+2}}{\p^2 \kappa^2 F\over \p t_k\p \tau_l}, \cr
&\chi^{BB}(z) =
\sum_{k,l\geq 0}{1\over z^{k+l+3}}{\p^2
\kappa^2 F\over \p t_k \p t_l}, \cr
&\chi^{FF}(z)=  \left(
\p_z\sum_{k\geq 0}{1\over z^{k+1/2}}{\p\over \p
\tau_k}\right)\sum_{l\geq 0}{1\over z^{l+1/2}}
{\p \kappa^2 F\over \p
\tau_l} \;\; .\cr}
\eqn\eiii
$$
The loop operators and the free energy have the following genus
expansion
$$
\eqalign{ &u(z)=u_0(z)+\k u_1(z)+\dots=\sum_{k\geq 0}
u_0^{(2k)}(z)+\k u_1^{(2k)}(z)+\dots ,\cr
&v(z)=v_0(z)+\k v_1(z)+\dots=\sum_{k\geq 0} v_0^{(2k+1)}(z)+\k
v_1^{(2k+1)}(z)+\dots ,\cr
&\chi^{(a)}(z)=\chi^{(a)}_0(z)+\k\chi^{(a)}_1(z)+\dots \qquad
a={BF,BB,FF} \;\; , \cr
& F=F_0+\k F_1+\dots \;\; .\cr}
\eqn\eiv
$$
The subindices in our notation indicate the genus, while for the
order in fermionic couplings we introduce upper indices.

{\it Planar solution}

The leading terms in the genus expansion of equations {\ei}
and {\eii} are
$$
\uo(z)\vo(z)={\rm Polynomial}(z)\;\; ,
\eqn\ev
$$
$$
\uo(z)^2-\vo(z)\p_z \vo(z)={\rm Polynomial}(z)\;\; .
\eqn\evi
$$
We follow closely the steps of the discrete case with a one-cut
ansatz for ${\widehat u}_0^{(0)}(z)$

$$
\uo(z)=M(z)\sqrt{\z}+{A(z)\over(\z)^{3/2}}\;\; ,
\eqn\evii
$$
$$
\vo(z)={N(z)\over \sqrt{\z}}\;\; .
\eqn\eviii
$$
Expanding in powers of $(\z)$
$$
N(z)=\sum_{k\geq 0}n_k(\z)^k \;\; ,
$$
$$
M(z)=\sum_{k\geq 0} m_k (\z)^k\;\; ,
\eqn\eix
$$
we see that $A(z)$ is determined by demanding the right hand
side of {\ev} to be polynomial in $z$
$$
\uo(z)=M(z)\sqrt{\z}-{1\over 2m_0}{n_1 n_0\over {(\z)}^{3/2}}\;\; .
\eqn\ex
$$
$M(z)$ is determined from $w^{(0)}_0\sim O(z^{-3/2})$. We have:
$$
{\p u_0^{(0)}\over \p u}={1\over \sqrt{\z}}({1\over
2}M(z)+(\z)\p_uM(z))-{1\over 2\sqrt{z}}{\p t_0\over \p u}\;\; ,
\eqn\exv
$$
which holds only when
$$
{1\over 2}M(z)+(\z)\p_u M(z)={1\over 2}{\p t_0\over \p u}\;\; .
\eqn\exvi
$$
Inserting {\eix} in this equation we obtain a relation between
the coefficients $m_k$ and the renormalized cosmological constant
$t_0$:
$$\eqalign{
& m_0={\p t_0\over \p u} \cr
&m_k={(-1)^k\over 2}{\Gamma({1\over 2}) \over \Gamma(k+{3\over
2})}\;{\p^{k+1}t_0\over \p u^{k+1}}
\qquad,\qquad k\geq 1 \;\; .\cr}
\eqn\exvii
$$
The modes of $N(z)$ are determined by demanding
$$
v_0(z)={N(z)\over \sqrt{\z}}-\eta(z)=O(z^{-1/2})\;\; .
\eqn\exii
$$
This implies ${\p v_0(z) \over \p u} \sim O(z^{-1/2})$,
leading to
$$
{\p n_k\over \p u}=-(k+{1\over 2})n_{k+1} \qquad k\geq 1\;\; .
\eqn\exiii
$$
Fermi-statistics and compatibility between the bosonic and
fermionic loop gives
$$
n_1=-4\p_u n_0\;\; .
$$
Thus
$$
\uo(z)=M(z)\sqrt{\z}-{2\over m_0}{n_0 \p_u n_0 \over(\z)^{3/2}}\;\; ,
\eqn\exviii
$$
$$
\p_t v_0(z)=-{1\over\sqrt{\z}}\DLR_t n_0\;\; .
\eqn\exix
$$
With the ansatz
$$
n_0=\sum_{n\geq 0}\beta_n u^n\;\; ,
$$
and {\exii} we obtain
$$
\beta_n={(-1)^n\Gamma(n+{1\over 2}) \over 2 n! \Gamma({1\over
2})}\tau_n\;\; .
$$
Thus
$$
n_0=-\tau(u)\qquad ,\qquad \tau(u)=\sum {\widetilde \tau_n}u^n\;\; .
\eqn\exiv
$$
The final result coming from {\exviii} and {\exix} is then
$$
\uo(z)=M(z)\sqrt{\z}+{\tau \p_t \tau \over (\z)^{3/2}}\;\; ,
$$
$$
\p_t v_0(z)={1\over \sqrt{\z}} \DLR \tau(u) \;\; ,
$$
and this coincides with the results of previous sections.

It is nice to see that from the purely bosonic part of $\uo(z)$
we obtain the planar string equation, since from:
$$
{u_0^{(0)}\over \sqrt{\z}}=M(z)-{t(z)\over \sqrt{\z}}=O(z^{-2})\;\;
,
\eqn\exx
$$
we obtain the expected result from the vanishing of the terms
proportional to $z^{-1}$:
$$
\sum_{k\geq 0}{(-1)^k\over
k!}{\Gamma(k+{3\over2})\over  \Gamma({1\over 2})} u^{k} t_k =0\;\; .
\eqn\exxi
$$

{\it Genus one solution}

The two superloop equations obtained from {\ei} and {\eii}
are\foot{
In the following we will omit the $z$ dependences}:
$$
2\uo u_1-v_1\p\vo-\vo\p v_1+
\chi_0^{BB}+\chi_0^{FF}+{1\over 4z^2}=
{\rm Polynomial}(z)\;\; ,
\eqn\exxii
$$
$$
v_1\uo+\vo u_1+\chi_0^{BF}={\rm Polynomial}(z)\;\; .
\eqn\exxiii
$$
${\widehat w}_0$ and ${\widehat v}_0$ are already determined.
The two-loop correlators are obtained from {\eiii} for $F=F_0$
$$\eqalign{
&\chi_0^{BF}=-{1\over2(\z)^{3/2}}\left(
{1\over\sqrt{\z}}\DLR_t \tau(u)\right) \;\; , \cr
&\chi_0^{BB}= {1\over 8}\left( {1\over
(\z)^2}-{1\over z^2} \right) -{1\over
2}\p_t{\tau\p_t\tau\over  (\z)^{3}} \;\; , \cr
&\chi_0^{FF}={1\over
8}\left({1\over (\z)^2}-{1\over z^2}\right)\;\; . \cr}
\eqn\exxiv
$$
We solve now the equations {\exxii} and {\exxiii} according
to the order of fermionic couplings
$$
u_1=\sum_{k\geq 0} u_1^{(2k)}\qquad v_1=\sum_{k\geq 0}
v_1^{(2k+1)}\;\; .
\eqn\exxv
$$

For the fermionic loop operator we obtain
$$
v_1=\sum_{k=0}^3 {V_{k+1/2}\over (\z)^{k+1/2}}\;\; ,
$$
with
$$\eqalign{
&V_{1/2}=-{1\over 3}D\left({ D^2 \tau  \over
Du}\right)\qquad\qquad \;\;\;\;\;\; \;\;\;\;\;\; V_{5/2}=
-{Du\DDLR \tau \over 2} \cr &V_{3/2}={2\over 3}D\left(
{V_{5/2}\over Du}\right) -\tau D\left({D^2u\over 2 Du}\right)
\qquad V_{7/2}= -{5\over 8}(Du)^2\tau\;\; .
\cr}
$$
Where we have introduced the notation $D=\p / \p t_0$.
The bosonic loop operator is
$$
u_1(z) = \sum_{k=1}^4 {W_{(k+1/2)}\over (\z)^{k+1/2}}\;\; ,
\eqn\exxvii
$$
with the coefficients
$$\eqalign{
&W_{3/2}={D^2u\over 12Du}-2\tau\DDLR V_{1/2}
\qquad\qquad\;\;\;\;\;\;\;\;\;\; W_{7/2}=5\tau D(Du D\tau)
\cr  &W_{5/2}=-{Du\over 8}+6V_{3/2}D\tau-2\tau \DDLR D^2 \tau
\qquad  W_{9/2}=7V_{7/2}\DDLR \tau\;\; . \cr}
\eqn\exxxiii
$$

The value of $V_{1/2}$ cannot be
determined by the requirement that the left hand side of
{\exxiii} is a polynomial in $z$.
It follows from the consistency between $v_1(z)$ and $u_1(z)$.
This condition is given by the requirement of commuting
derivatives
$$
{\p^2 F_1\over \p t_0\p \tau_k}={\p^2 F_1 \over \p \tau_k \p
t_0}
\eqn\exxxiiibis
$$
where $F_1$ denotes the genus one contribution to the free
energy.

For the purely bosonic part of $u_1(z)$ (no presence of
fermionic couplings) one easily sees that this is the same
result as for the one-matrix model.
The equation to be solved in this case is
$$
{\widehat u}_0(z)u_1(z)+\chi_0(z)+{1\over 8z^2}={\rm
Polynomial}(z)\;\; ,
\eqn\exxxiv
$$
$$
\chi_0(z)={1\over 8}\left( {1\over (\z)^2}-{1\over z^2}\right)\;\; ,
\eqn\exxxv
$$
where ${\widehat u}_0(z)=M(z)\sqrt{\z}$ is the solution for
genus zero. Equation {\exxxiv} coincides with the order
zero equation in fermionic couplings coming from {\exxii} just
because the bosonic two-point correlators {\exxiv} and
{\exxxv} coincide in the bosonic part. Our result does not agree
with the continuum Liouville calculation for genus one of
[\berkleb]. The origin of this discrepancy may be the fact,
that there was no simple way to normalize the supermoduli
integration in [\berkleb].

The reader may have noticed, that the solution presented for
$v_1(z)$ is only linear in the fermionic couplings, as it was
for genus zero. This comes from the fact that, the third order
equation in fermionic couplings
$$
v_1^{(3)} {\widehat u}_0^{(0)}+{\widehat v}_0u_1^{(2)}+v_1^{(1)}
{\tau \p_t \tau
\over \D^{3/2}} ={\rm Polynomial}(z)
\eqn\exxxvi
$$
is solved by $v_1^{(3)}=0$ which means that also
on the torus we have a maximal coupling of two fermions.

{\it Genus two solution}

We consider now the situation in genus two to see if the free
energy changes with respect to the one-matrix model. The two
superloop equations to be solved are
$$
2{\widehat u_0}u_2+u_1^2-v_2\p {\widehat v_0} -{\widehat v_0}\p v_2
-v_1\p v_1 +\chi_1^{BB}+\chi_1^{FF}={\rm Polynomial}(z)
\;\; ,
\eqn\exxxvii
$$
$$
{\widehat v_0}u_2+{\widehat u_0}v_2+u_1 v_1+\chi_1^{BF}=
{\rm Polynomial}(z)\;\; .
\eqn\exxxviii
$$
To calculate the free energy, we are only interested in
the purely bosonic piece of the two loop correlators, which
are given by (take {\eiii} for $F=F_1$)
$$
\chi_1^{\rm FF}={\chi_1^{\rm BB}}^{(0)}=
{13\over 16 m_0^2}{1\over \D^5}-{3\over 4}{m_1\over
m_0^3}{1\over \D^4}+\left({3\over 8}{m_1^2\over
m_0^4}-{5\over 16}{m_2\over m_0^3}\right){1\over \D^3} \;\;.
\eqn\exli
$$
We see, that again these two loop correlators coincide with
the one-matrix model values. This immediately implies, that
also in genus two we have the same value in the purely
bosonic part of the free energy as in the one-matrix model.
The complete solution to order zero in fermionic couplings is
$$\eqalign{
&u_2^{(0)}={a \over \D^{3/2}}+{b\over \D^{5/2}}+{c\over
\D^{7/2}}+{d\over \D^{9/2}}+{e\over \D^{11/2}}\cr
&a=-{63 m_1^4\over 32 m_0^7}+{75
m_1^2 m_2\over 16 m_0^6}-{145 m_2^2 \over 128 m_0^5}-
{77 m_1m_3\over 32
m_0^5}+{105 m_4 \over 128 m_0^4}\cr
&b={63 m_1^3 \over 32 m_0^6}-{87 m_1 m_2
\over 32 m_0^5}+{105
m_3 \over 128 m_0^4}\cr
&c=-{63 m_1^2 \over 32 m_0^5}+
{145 m_2 \over 128 m_0^4}\qquad d={203
m_1 \over 128 m_0^4}\qquad e=-
{105 \over 128 m_0^3}\;\;.\cr}
\eqn\exlii
$$
We now continue the calculation for the case of pure gravity
($m_k=0,\;\;k\geq 2$). Our results reproduce, up to genus two the
values expected for the Painlev\'e-I equation. Since
$$
m_0={1\over Du} \qquad m_1={2\over 3}m_0^3D^2 u
\eqn\exliii
$$
we get for the second derivative of the free energy the expansion
$$
\langle \sigma_0 \sigma_0 \rangle=
-{u\over 4}+{\k\over 12}D\left({D^2 u \over
Du}\right)-\kappa^4 {63\over 162}D\left((D^2u)^4\over
(Du)^5\right)+\dots \;\; .
\eqn\exliv
$$
For pure gravity we take $u=\sqrt{t}$, to obtain
$$
\langle \sigma_0 \sigma_0 \rangle=-{1\over 4}
(\sqrt{t}-{1\over 24}{\k \over t^2}-{49 \over 1152}{\kappa^4
\over t^{9/2}}+\dots )
\eqn\exlv
$$
which agrees with the first three terms of the solution to the
Painlev\'e-I equation appearing in pure gravity [\doubles].

{\it General properties of the Free Energy}

We now give the heuristic argument showing that the bosonic
loop operator $u(z)$ coincides with the pure gravity case
with an even potential
when all the fermion couplings are
set to zero.  First we write the loop equations
for the pure gravity case derived from an even potential.
They are given by
$$
\h \u^2 +\kappa^2(\chi(z)+{1\over 8z^2})={\rm Polynomial}(z).
\eqn\nkdvi
$$
Note that equation (6.29) implies $\chi(z)+{1\over 8z^2}
={1\over 8}\D^{-2}$, where $\D=z+u$.  The correlator
$\chi(z)$ can be written in terms of the loop creation
operator as in the supersymmetric case
$$
\chi(z)={\cal J}_F (z)u(z).
\eqn\nkdvii
$$
The explicit form of the loop insertion operators
can be found in (4.41),(5.28).  Since $\uo=M(z)\D^{\h}$,
$u_n(z)$ for $n>0$ will be obtained in terms of
the negative half-integer powers of $\D$,
$$
u_n(z)=\sum_{k\ge 1}u_n^k\D^{-k-1/2}.
\eqn\nkdviii
$$
The expansion in \nkdviii\ starts with $\D^{-3/2}$
because of the general structure of the loop operator
discussed previously.  Similarly, by (4.34)
${\cal J}_B\sim O(\D^{-3/2})$, and the first power
of $\D$ in $\chi(z)$ is $\D^{-3}$.  These general
remarks about the power series representation
of $\chi(z0,u(z)$ apply beyond genus zero.  On
the sphere there are non-universal contributions
like $1/z^2$ in (6.29) and we do not have such
simple expansions in $\D$.  These considerations
will be important in the following.

The supersymmetric loop equations are given by
(6.1,2).  Since we are only interested in the
leading contribution $u^{(0)}(z)$ i.e. to order
zero in fermionic couplings, it suffices to
keep $\chi^{BB}$ and $\chi^{FF}$ to order zero
and $v(z)$ to first order in fermionic couplings.
Similarly, we can drop the term $v\partial v$
in (6.2) which becomes,
$$
{\widehat u}(z)^2+\kappa^2(\chi^{BB}(z)+\chi^{FF}(z)
+{1\over 4 z^2})={\rm Polynomial}(z).
\eqn\nkdviv
$$
We showed by direct computation that $\chi^{BB}_n(z)=
\chi^{FF}_n(z)$ at least for $n=0,1$, the sphere and
the torus. If this were true to all orders, then
\nkdviv\ would become:
$$
\h {\widehat u}(z)^2+\kappa^2(\chi^{BB}(z)+{1\over 8 z^2})=
{\rm Polynomial}(z),
\eqn\nkdvv
$$
which is identical to \nkdvi, and our proposition
would be proved.  We will argue by induction
that indeed  $\chi^{BB}_n(z)= \chi^{FF}_n(z)$.  Assume
this to be true up to order $n-1$.    Expand \nkdviv\
in powers of $\kappa^2$.  This yields,
$$
2\uo u_n+\sum_{k=1}^{n-1} u_k u_{n-k}
+2\chi^{BB}_{n-1}={\cal O}(1),
\eqn\nkdvvi
$$
where we have used  $\chi^{BB}_{n-1}(z)=
\chi^{FF}_{n-1}(z)$, and we have labeled a polynomial
in $z$ as ${\cal O}(1)$ in $\D$.  We have included
the non-universal factor $t(z)$ in $\uo$.  Expanding
(6.2) in powers of $\kappa^2$ we obtain
$$
\uo v_n+u_n\vo +\sum_{k=1}^{n-1}u_k v_{n-k}
+\chi^{BF}_{n-1}={\cal O}(1).
\eqn\nkdvvii
$$
Acting with ${\cal J}_B$ on \nkdvvi\ yields
$$
2\uo \chi^{BB}_n +2\chi^{BB}_0 u_n +2\sum_{k=1}^{n-1}
u_{n-k}\chi^{BB}_{n-k} +2{\cal J}_B \chi^{BB}_{n-1}=
{\cal O}(\D^{-3/2}),
\eqn\nkdvviii
$$
next, acting with $\partial_z {\cal J}_F$ on \nkdvvii\
we obtain
$$
\uo \chi^{FF}_n(z) +u_n {\widehat \chi}^{BB}_0
+\sum_{k=1}^{n-1}u_k\chi^{BB}_{n-k}+
\partial_z {\cal J}_F\chi^{BF}_{n-1}=
{\cal O}(\D^{-3/2}),
\eqn\nkdvix
$$
where we have used the induction hypothesis.  For the
same reason, we have the equalities
$$
\partial_z {\cal J}_F\chi^{BF}_{n-1}=
\partial_z {\cal J}{\cal J}_B v_{n-1}=
{\cal J}_B \chi^{FF}_{n-1}=
{\cal J}_B \chi^{BB}_{n-1},
\eqn\nkdvx
$$
and we find that \nkdvviii\ and \nkdvix\ are identical.  Since
$\uo={\cal O}(\D^{-1/2})$, this would imply that
$\chi^{BB}_n=\chi^{FF}_n$,  as was to be shown.  The basic
subtlety in this argument has to do with the
equality between ${\widehat \chi}^{BB}_0$
and  ${\widehat \chi}^{FF}_0$.  This is because naively
these quantities contain infinities.  To regulate them we
have to point-split the variable in the loop creation
operator with respect to the variable $z$ in which
we express the loop equations.  This makes the
count of powers of $\D$ rather subtle, and to assure
that no mistakes are made we have to study in detail the
two-loop equations.  Nevertheless, we believe that
after the appropriate subtleties of the mathematical
details are clarified, the conclusion of this heuristic
proof will remain true.

\chapter{RELATION TO SUPERINTEGRABLE HIERARCHIES}

Up to this point we have been able to calculate correlation
functions for the first cases in the genus expansion.
Although this method is straightforward to apply for any
genus, the procedure is cumbersome and of course it is
essential to find the relation to supersymmetric integrable
hierarchies.

This relation is well  established for the one-matrix model
\REF\lagcgl{L. Alvarez-Gaum\'e, C. Gomez
and J. Lacki {\it Phys. Lett. }{\bf B253} (1991)
56}[\douglas, \doubles, \lagcgl]  . If we denote the specific heat
with $U=\langle PP \rangle $ where $P$ is the puncture operator,
then the different multicritical points are characterized by the
string equation
$$
-x=\sum_{n=1}^\infty t_n R_n [U],
\eqn\gi
$$
here $R_n[U]$ are the Gelfand-Dikii polynomials of the
KdV-hierarchy, defined through the recursion relations
$$
DR_{n+1}[U]=(\k D^3+4UD+2(DU))R_n[U]
\eqn\gii
$$
$R_0={1\over 2}$, $R_1=U$, $R_2=(3U^2+\k U'')$, $\dots$

The renormalization group flows of the models described by
{\gi} coincide with the flows of the KdV-hierarchy:
$$
{\p U\over \p t_n}=DR_{n+1}[U]
\eqn\giii
$$
{\gii} implies therefore recursion relations between the
flows. Equations {\gi} and {\gii} contain all the
information about the correlation functions of the model.

Through the recent work of Witten and Kontsevich [\witkon] we can
directly relate the Virasoro constraints $L_n\tau=0,\;\; n\geq -1$
with
 the
$\tau-$function for the KdV hierarchy with initial condition
$L_{-1}\tau =0$. Since in the supercase we do not yet have a
formulation in terms of generalized matrices, it is useful to take
a more pedestrian approach and see how the KdV-hierarchy emerges
from the explicit solution of the loop equations. In the purely
bosonic model, the one-point functions are given by:
$$
D\langle \sigma_n\rangle _0={(-1)^{n+1}\over 2 n!}{\Gamma(n+{3\over
2}) \over \Gamma({1\over 2})}{u^{n+1}\over n+1}
\eqn\gv
$$
The definition of $U$ as the two-point function of the puncture
operator implies:
$$
D \langle
\sigma_0 \rangle_0=U^{(0)}=-{1\over 4}u
\eqn\gvi
$$
Using the planar string equation and $\gvi$ we learn that
$$
D^2\langle \sigma_n \rangle ={\p U^{(0)}\over \p t_n}= DR_{n+1}^{(0)}
\quad , \quad
R_{n+1}^{(0)}={2^{2n+1}\Gamma(n+{3\over 2}) \over (n+1)! \Gamma
({1\over
2})}U^{n+1}
\eqn\gvii
$$
At genus one,
$$
w^{(1)}(z)={m_1\over 8m_0^2}{1\over (\z)^{3/2}}-{1\over 8m_0}{1\over
(\z)^{5/2}}
\eqn\gviii
$$
Therefore,
$$
\langle \sigma_n \rangle_1=\k {(-1)^n \Gamma(n+{3\over 2}) \over
2m_0  n! \Gamma({1\over 2})} \left( {m_1\over 2m_0}u^n+{n \over
3}u^{n-1}\right)
\eqn\gix
$$
Hence the ``heat capacity'' $U$ is given to genus one by:
$$
U=U^{(0)}+\k U^{(1)}=-{1\over 4}u+{1\over 12}\k D\left({D^2 u\over
Du}\right)
\eqn\gx
$$
Similarly, to this order we can compute $\p U/ \p t_1$, and
obtain the well known KdV equation
$$
{\p U \over \p t_1}=\k D^3 U+3DU^2
\eqn\gxi
$$
Hence, to this order:
$$
R_2[U]=\k D^2 U+3U^2,
\eqn\gxii
$$
and
$$
R_{n+1}[U]={2^{2n+1}\Gamma (n+{3\over 2}) \over n!
\Gamma({1\over 2})}
\left({U^{n+1}\over n+1}+{n \over 12}\k U^{n-1}D^2 U+{\k \over 12}D^2
U^n\right) +O(\kappa^4)
\eqn\gxiii
$$
Including now the genus two correction to $U$ which was computed in
the previous section:
$$
U=U^{(0)}+\k U^{(1)}+\kappa^4U^{(2)}
\eqn\gxiv
$$
it is easy to see that at genus two there is no correction to {\gxi},
and furthermore, that to this order we have agreement with {\gii}.
Although this is by no means the cleanest way to exhibit
the equivalence between
KdV and the loop equations, it is clear that to determine the
differential operator appearing on the right hand side of {\gii}
(assuming its existence) we need to know explicit correlators only
for genera zero, one and two. Equation {\gxi} together with the
fact that it is not corrected in genus two and the perturbative
understanding of {\gii} provide strong hints that the flows in the
one-matrix model are governed by the KdV-hierarchy.
In the supesymmetric case, we could proceed by analogy with the
previous arguments. Since we have computed all correlation functions
for genus zero and one, and we have partial results in genus two, we
can try to explore what type of differential relations allow us to
express the flows $\p / \p t_n$, $\p / \p \tau_n$ in terms of the
basic flows $\p /\p t_0$, $\p / \p \tau_0$. We can also examine the
supersymmetric extensions of the KdV- or KP-hierarchies. For KdV,
the papers [\skdv] find a one parameter family of supersymmetric
extensions given by:
$$
{\dot u}=-u'''+6uu'+3ia \xi \xi'' ,
\eqn\gxv
$$
$$
{\dot \xi}=-a\xi '''+(2+a)u \xi '+3u' \xi ,
\eqn\gxvi
$$
$$
{\dot f}={\p f \over \p t} \qquad f' ={\p f \over \p x}
$$
$u$ (resp. $\xi$) is a bosonic (resp. fermionic) function. This
system
is integrable only for $a=1,4$. Only the case $a=1$ is really
supersymmetric. If we formally set $\xi=0$, we obtain KdV.
Equations {\gxv} and {\gxvi} can be therefore considered as
supersymmetric extensions to the KdV equation. If on the other
hand, we are also interested in fermionic flows ($\p / \p \tau_n$),
we need to extend these systems to incorporate odd flows. This has
been done in [\mr, \mulase, \rabin]. In reference [\mr] the
KP-hierarchy is extended using a supersymmetric Lax-pair
formulation. A different approach is taken in [\mulase, \rabin],
where only the even (i.e. bosonic) flows admit a Lax pair
representation. The fermionic flows have a simpler expression than
those of [\mr]. A common feature of all these supersymmetric
hierarchies is that if we consider only even flows, and we set all
the fermionic variables to zero, we recover the standard KdV- or
KP-hierarchies. In terms of loop equations this is similar to
formally considering only the bosonic part of the energy momentum
tensor $T_B(z)$ and ignore the fermion field dependence. This would
generate the KdV-hierarchy as in the bosonic case. In this respect
our loop equations are a good starting point to construct another
supersymmetric extension of KdV.

If we take the fermionic dependence into account, we have not yet
been able to identify the corresponding differential relations
which should lead to a local integrable hierarchy. Our analysis
does not seem to make our one- or higher-point functions compatible
with known extensions {\gxv} and {\gxvi} or those in [\skdv, \mulase,
\rabin]. This is perhaps not so bad as it sounds. If our model indeed
describes a non-critical superstring, the alleged hierarchy
describing
the continuum limit should capture the geometry of the supermoduli
space
of super-Riemann surfaces. From the work in [\rabin] we know that the
algebro-geometric data needed to construct quasi-periodic solutions
to the Manin-Radul or Mulase-Rabin hierarchies does not include
super-Riemann surfaces. We continue looking for an integrable
supersymmetric hierarchy compatible with our loop equations.

\chapter{CONCLUSIONS AND OUTLOOK}
In this paper we have taken the double scaling limit of the superloop
equations proposed in [\sloops]. Working with general potentials, we
have shown that the spectrum of anomalous dimensions coincides with
those which follow from [\pz,\dhk],
when one couples two-dimensional
supergravity to minimal superconformal matter of type $(2,4m)$. The
continuum limit of these theories is described by a $Z_2$-twisted
scalar representing bosonic loops, and a Weyl-Majorana fermion in the
Ramond sector representing fermionic loops. We have solved the
continuum superloop equations in genus zero and one completely, and
partially in genus two. The piece of the free energy independent of
the fermionic couplings is
given in the pure supergravity case by the
same expression as for the one-matrix models
[\doubles] i.e. by the same
solution to the Painlev\'e-I equation. This conclusion was shown
to hold in general by a heuristic argument in Section seven.
So far we have not been able to identify an integrable
superhierarchy,
which reproduces our correlation functions. As explained in the text,
this is not necessarily negative. No superhierarchy is yet known,
which
incorporates fully the geometry of super-Riemann surfaces. Finally,
the corresponding generalization of multimatrix models in our
context is still missing. Work in these directions is in progress,
and we hope to report on the results elsewhere.

\vskip 1cm

{\bf ACKNOWLEDGEMENTS.}
We have had useful
discussions of integrable superhierarchies with M. Mulase and J.
Rabin. We are grateful to them for explaining to us the state of
this field. We are grateful to H. Itoyama and A. Zadra for
discussions during the course of this work. K.B and M.B. would
like to thank W. Nahm for his help and encouragement and the Theory
Division of CERN for hospitality.  R.E. was supported in part
by a FPI scholarship from the Ministerio de Educaci\'on y
Ciencia, Spain.  J.L.M. was supported in part by a
CICYT grant AEN90-0330 and by a UPV research grant
UPV172.310-E035/90.

{\bf NOTE ADDED.} This work supersedes the preprint ``Superloop
Equations in the Double Scaling Limit'' CERN-TH.6575/92 by three of
the authors (L.A.G., K.B. and M.B.)

\REF\kpz{\rkpz}
\REF\matmod{\rmatmod}
\REF\pz{\rpz}
\REF\migdal{\rmigdal}\REF\mmreview{\rmmreview}
\endpage
\refout
\end